\documentclass[twocolumn,showpacs,superscriptaddress,preprintnumbers,amsmath,amssymb]{revtex4-2}
\usepackage{graphicx}
\usepackage{dcolumn}
\usepackage{appendix}
\usepackage{bm}

\usepackage[usenames,dvipsnames]{xcolor}
\usepackage{subfigure}
\usepackage{bbm}
\usepackage{enumerate}
\usepackage[
  bookmarks=true,
  colorlinks,
  linkcolor=blue,
  urlcolor=blue,
  citecolor=blue,
  plainpages=false,
  pdfpagelabels,
  final,
  breaklinks=true
]{hyperref}

\usepackage{titlesec}
\usepackage{array}
\usepackage{dsfont}
\usepackage{physics}
\usepackage{mathpazo}

\begin{document}

\title{Attosecond metrology of bright quantum light}

\author{P.~Stammer}
\email{philipp.stammer@icfo.eu}
\affiliation{ICFO -- Institut de Ciencies Fotoniques, The Barcelona Institute of Science and Technology, 08860 Castelldefels (Barcelona), Spain}
\affiliation{Atominstitut, Technische Universit\"{a}t Wien, 1020 Vienna, Austria}

\author{J.~Rivera-Dean}
\affiliation{Department of Physics and Astronomy, University College London, Gower Street, London WC1E 6BT, United Kingdom}

\author{E.~Pisanty}
\affiliation{Attosecond Quantum Physics Laboratory, Department of Physics, King’s College London, Strand Campus, London WC2R 2LS, UK}

\author{M.~Lewenstein}
\affiliation{ICFO -- Institut de Ciencies Fotoniques, The Barcelona Institute of Science and Technology, 08860 Castelldefels (Barcelona), Spain}

\date{\today}

\begin{abstract}

Attosecond metrology is the ability to measure ultrafast optical light-wave oscillations, yet its approach has been limited to classical fields. Hence, the influence of the fluctuations of a quantum field on attosecond measurements has remained unexplored. Here, we close this gap by showing that the attosecond streaking measurement of bright quantum light is sensitive to quantum fluctuations of the optical field on the attosecond timescale. The distinct sub-cycle modulations allow to extract the properties of the squeezed field quadrature in regimes where conventional state tomography approaches reach their limitation. With the full quantum optical attosecond streaking scheme developed here, we provide a certification method that can measure quantum squeezing below the shot noise limit, thereby overcoming the problem of tomographically measuring bright quantum light. This opens the way towards quantum optical metrology of field fluctuations with attosecond temporal resolution.

\end{abstract}

\maketitle

\textit{Introduction.--}
Attosecond streaking is the ability to perform ultrafast metrology of optical light-wave oscillations and to reconstruct isolated attosecond pulses~\cite{hentschel2001attosecond, drescher2001x, kienberger2004atomic}.
In essence, the technique of attosecond streaking maps the temporal profile of an attosecond burst of light onto the distribution of photoelectron energies~\cite{krausz2009attosecond}, which encodes the information for the retrieval of the pulse~\cite{mairesse2005frequency, quere2005temporal}.  

The basic principle underlying attosecond streaking is the ionization of a photoelectron by a short extreme-ultraviolet (XUV) attosecond burst, in which the electron interacts with an infrared (IR) streaking field. The streaking field drives the electron in the continuum, and measuring its momentum distribution encodes information about the ionization time~\cite{schultze2010delay}, the characteristics of the XUV pulse~\cite{drescher2001x, itatani2002attosecond, kitzler2002quantum}, and about the IR field~\cite{goulielmakis2004direct}.
Classically, the streaking field adds a momentum shift to the electron at the instant of ionization via the XUV pulse~\cite{krausz2009attosecond, itatani2002attosecond}, allowing for the measurement of the optical light-wave oscillations~\cite{goulielmakis2004direct}. From the semi-classical perspective~\cite{kitzler2002quantum, itatani2002attosecond}, in which the quantum state of the electron describes a wavepacket in the continuum, the distribution encodes the information of the oscillating laser field acting as an attosecond phase gate on the wavepacket released by the XUV pulse.
In fact, the semi-classical theory of attosecond streaking has led to a tremendous progress in attosecond science~\cite{krausz2009attosecond, corkum2007attosecond, Huillier_Nobel_2024, Agostini_Nobel_2024, Krausz_Nobel_2024}. 

However, the existing description of attosecond streaking is limited to classical fields, and the quantum nature of light has thus far been ignored~\cite{thesis_stammer}. On the other hand, quantum optics studies the interactions of individual photons with matter, with the quantum nature encoded in the fluctuations and correlations of the field. To retrieve this information there exists a variety of advanced experimental techniques for measuring the field fluctuations, such as homodyne detection for measuring the field quadratures~\cite{wiseman1993quantum, breitenbach1997homodyne, leonhardt1995measuring, lvovsky2009continuous}, and to reconstruct phase-space distributions~\cite{schleich2015quantum}. 
Of particular interest are squeezed states of light due to their wide applicability in optical quantum technologies~\cite{dowling2015quantum, dowling2003quantum}. 
This includes, for instance, the ability to perform ultrasensitive measurements for the detection of gravitational waves~\cite{caves1981quantum, schnabel2010quantum} or quantum sensing~\cite{lawrie2019quantum}, and is particularly useful for continuous-variable quantum information processing~\cite{weedbrook2012gaussian, braunstein2005quantum} or quantum communication~\cite{usenko2026continuous}. 
The usefulness of squeezed light, compared to classical coherent laser light, originates from the significantly different quantum fluctuations, where one of the field quadratures exhibits fluctuations below the shot noise vacuum limit~\cite{Walls1983, breitenbach1997measurement}. 
Yet, the existing schemes to tomographically measure genuine quantum squeezing below the shot noise limit are restricted to moderate field intensities. This is due to the inherent scaling problem between the signal and reference field and limited by practical means due to detector efficiencies~\cite{raymer20047, tiedau2019high, knyazev2018quantum, yoon2026efficient}.
Therefore, the current certification approaches of bright quantum light sources usually rely on the measurement of the equal time intensity correlation function $g^{(2)}(0)$. The observation of a super-Poissonian photon number distributions, with its typical long tail in the intensity distribution, is usually associated to squeezing~\cite{breitenbach1997measurement}. Yet, these measurements can not certify quantum squeezing since there exist classical counterparts having the same distribution. In particular when the field is in a multi-mode squeezed state, the $g^{(2)}(0)$ measurement can show similar values as a thermal state. This results in the imminent need for a solution to detect and certify bright squeezed light.

\begin{figure}
    \centering
	\includegraphics[width=0.7\columnwidth]{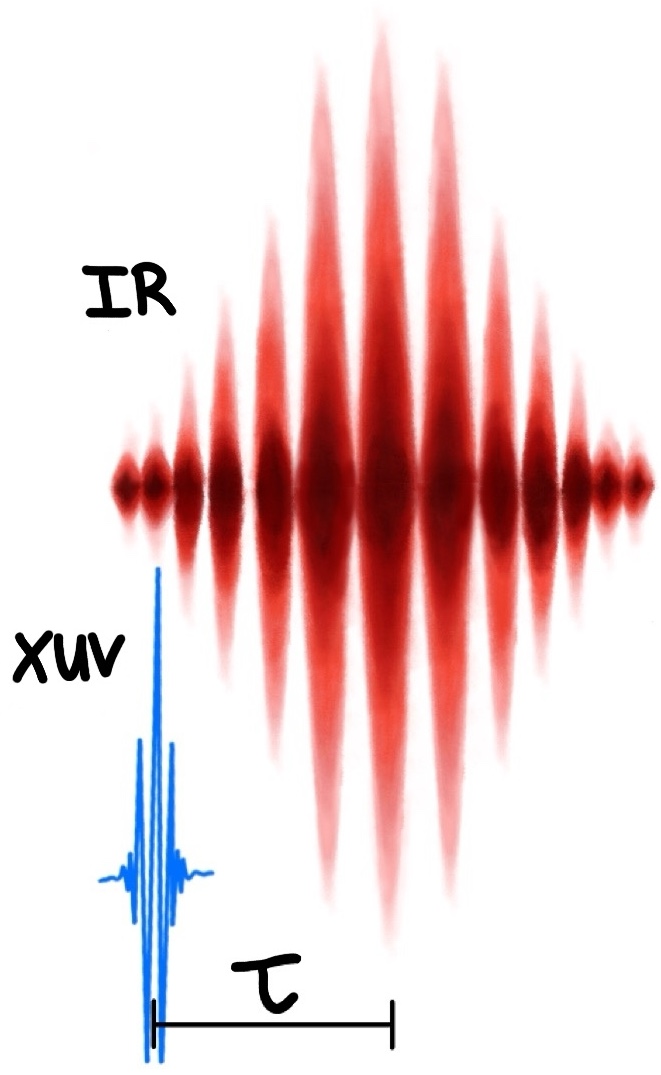}
	\caption{\textbf{Streaking of a bright squeezed vacuum field.} Schematic illustration of the quantum optical attosecond streaking, where the IR streaking field consists of quantum light such as a bright squeezed vacuum and is delayed by $\tau$ with respect to the coherent XUV pulse. The distinct subcycle modulations in the streaking trace reveal the field fluctuations and allow for measurement of quadrature squeezing.}
      \label{fig:cartoon}
\end{figure}

In this work we overcome this limitation and introduce a new approach to measure the quantum fluctuations of light on the attosecond time-scale.
This is possible due to the recent advances in strong-field quantum optics~\cite{stammer2025colloquium}, and particularly the development of bright squeezed light sources~\cite{spasibko2017multiphoton, rasputnyi2024high, heimerl_multiphoton_2024, heimerl2025quantum, lemieux2024photon, kern2026single}, challenging the semi-classical perspective of strong-field phenomena~\cite{stammer2024limitations, cruz2024quantum, rivera2025structured, stammer2026fluctuation}. Of special interest is the process of high harmonic generation (HHG), in which a strong IR field generates a frequency comb leading to the emission of attosecond XUV bursts~\cite{lewenstein1994theory, antoine1996attosecond, krausz2009attosecond}. 
While the quantum optical understanding of HHG has led to new insights about the properties of the harmonics~\cite{lewenstein2021generation, gorlach2020quantum, stammer2024entanglement, yi2024generation, lange2024electron, lange2024hierarchy, stammer2022high, gombkotHo2021quantum, stammer2025theory, wang2025high, pizzi2023light, stammer2026photon, lange_excitonic_2025, de2024quantum, boroumand2025quantum}, the applications towards quantum optical measurement schemes and techniques is thus far limited~\cite{rivera2026attosecond, stammer2025weak, lamprou2025nonlinear, dubois2026quantum}. In particular, the imminent challenge of quantum state tomography of bright quantum states of light~\cite{mele2025learning, rasputnyi2026kerr}, in which conventional methods of homodyne detection reach their limitations, is yet unresolved. This leads to a gap in certifying the squeezing properties of intense quantum fields of light.
The scheme introduced here overcomes this problem, by using the attosecond streaking technique to certify the presence of squeezing in bright quantum light (see Fig.~\ref{fig:cartoon}).

\textit{Quantum optical theory of attosecond metrology.--}
In the following we present the formalism to describe attosecond streaking measurements of an arbitrary light field. The light field can be in a classical state, for instance described by the coherent state $\ket{\alpha}$, or quantum state such as a bright squeezed vacuum $\ket{\xi}$. 
To this end, we derive analytical expressions for the delay-dependent attosecond streaking trace, revealing the ability to measure quantum fluctuations of light on the attosecond time-scale.

Before deriving the general expressions, it is intuitive to revisit the streaking measurement for a classical coherent field~\cite{yakovlev2005attosecond, kitzler2002quantum, yakovlev2010attosecond}.  
Therefore, we consider the interaction of a linearly polarized XUV pulse with the electonic system in the initial stationary bound state $\ket{\psi(t_0)}$, and the pulse generates a single-electron wavepacket $\ket{\psi(t)}$ due to photoionization. The classical XUV pulse has the electric field $\vb{E}_X(t) = \operatorname{Re}[\mathcal{E}_X(t) e^{- i \Omega t}]$, where $\mathcal{E}_X(t)$ is the complex envelope with central frequency $\Omega$. 
The electron wavepacket launched into the continuum will be detected with asymptotic momentum $\vb{p}$ with probability $\abs{\bra{\vb{p}}\ket{\psi(t)}}^2$, such that the photoelectron spectrum in the absence of a streaking field is given by $S_0(\vb{p}) = \abs{\bra{\vb{p}}\ket{\psi(t)}}^2$. Now, in the presence of a classical coherent laser field $\vb{E}_\alpha(t) =\bra{\alpha} E_Q(t) \ket{\alpha} = \operatorname{Re}[\mathcal{E}_{\alpha} (t) e^{- i \omega t}]$, which is delayed by $\tau$ with respect to the ionizing XUV pulse, the spectrogram, $S_\alpha(\vb{p}, \tau) = \abs{M_\alpha(\vb{p}, \tau)}^2$, can be written in terms of the streaking amplitude (atomic units used throughout)
\begin{align}
    M_\alpha(\vb{p}, \tau) = \int dt \, \mathcal{E}_X(t+\tau) G_\alpha(\vb{p}, t) e^{i \Delta_p t},
\end{align}
where $G_\alpha(\vb{p}, t) = D(\vb{p}+ \vb{A}_\alpha(t)) \exp [i \Phi_\alpha(\vb{p},t)]$ is the IR dependent gate function with the transition dipole moment $D(\vb{k}(t)) = \bra{\vb{k}(t)} \vb{d} \ket{\psi(t_0)}$, and the Volkov phase 
\begin{align}
    \Phi_\alpha(\vb{p},t) = - \int_t^\infty dt' \left( \vb{p}\cdot \vb{A}_\alpha(t') + \frac{1}{2} \vb{A}_\alpha^2(t') \right),
\end{align}
with the classical vector potential $\vb{A}_\alpha(t)$ corresponding to the coherent field $\vb{E}_\alpha(t) = - \partial \vb{A}_\alpha(t) / \partial t$. We have further defined the energy difference $\Delta_p = (\abs{\vb{p}}^2 - \abs{\vb{p}_0}^2)/2$, between the measured electron with momentum $\vb{p}$ and initial momentum $\lvert \vb{p}_0 \rvert = \sqrt{2(\Omega + E_0)}$ after the single photon ionization from the stationary ground state with energy $E_0 < 0$.  

Now, instead of the interaction with the classical laser $\vb{E}_\alpha(t)$, we consider the coupling of the system to the quantized electromagnetic field~\cite{stammer2023quantum}.
Therefore, we consider the dipole coupling of the electron to the field $H_I(t) = - \vb{d} \cdot \vb{E}_Q(t)$, where the time-dependent electric field operator includes the relevant modes of the streaking process
\begin{align}
    \vb{E}_Q(t) = -i \left( g_0  a^\dagger  e^{i \omega t} + g  b^\dagger  e^{i \Omega t} \right) + \operatorname{h.c.},
\end{align}
where $a^{\dagger}$ and $b^\dagger$ are the creation operators of the IR and XUV field, with frequencies $\omega$ and $\Omega$ as well as coupling constants $g_0$ and $g$, respectively.
The field operator is given in the rotating frame with respect to the free-field Hamiltonian $H_F = \omega a^\dagger a + \Omega b^\dagger b$.
Going beyond the existing semi-classical descriptions of attosecond streaking~\cite{thesis_stammer}, we further need to define the quantum states of the relevant fields. For the XUV pulse, we consider the field to be in a coherent state $\ket{\beta_X}$, while the IR streaking field is in an arbitrary quantum state $\rho_{IR}$. Note that the coherent state of the XUV pulse corresponds to the mode described by excitations of the field operator $b^\dagger = \int dt \phi_b(t) \mathcal{E}_X^\dagger(t)$, with the temporal mode $\phi_b(t)$ corresponding to the XUV pulse.
The continuous mode operators $\mathcal{E}_X^\dagger(t)$ can be understood as the excitations corresponding to the total XUV photons $N_X = \expval*{\mathcal{E}_X^\dagger(t) \mathcal{E}_X(t)} = \sum_i \bar n_i \phi_i^*(t) \phi_i(t)$ for the orthonormal set of temporal modes $\phi_i(t)$ with corresponding mean photon occupation number $\bar n_i$ in each mode~\cite{fabre2020modes}. 
To obtain the quantum optical streaking trace (QOST) we are essentially interested in the probability of finding the electron with asymptotic momentum $\vb{p}$, when driven by an arbitrary quantum state of the IR streaking field. Taking into account the quantum state of the light field, the QOST is obtained from 
\begin{align}
\label{eq:streaking_trace}
    S_Q(\vb{p}, \tau) = \Tr \left[ \dyad{\vb{p}}  \rho_\infty(\tau) \right],
\end{align}
where the trace is taken over the electron and field degrees of freedom with the projection on the final electron momentum $\dyad{\vb{p}}$ for the electron measurement. We have further defined the joint light-matter quantum state
\begin{equation}
\begin{aligned}
    \rho_\infty(\tau) & = \lim_{t \to \infty} U(t;\tau) \rho(0) U^\dagger(t;\tau) \\
    & = U_\infty(\tau) \rho(0) U^\dagger_\infty(\tau),
\end{aligned}
\end{equation}
which parametrically depends on the XUV-IR delay $\tau$, and gives the joint state of the asymptotic evolution under the unitary $U(t;\tau)$ containing the interaction. The initial state of the electron plus field is given by 
\begin{align}
    \rho(0) = \dyad{\psi(t_0)} \otimes \rho_F(t_0),
\end{align}
where the electronic state $\ket{\psi(t_0)}$ is the state from the known semi-classical picture, and we have extended the approach by introducing the initial quantum state of the field $\rho_F(t_0)$. 
An instrumental example is given in the particular case of a pure initial field state, such that $\rho_F(t_0) = \dyad{\Phi(t_0)}$, and hence the general expression of the streaking trace in Eq.~\eqref{eq:streaking_trace} simplifies to 
\begin{align}
\label{eq:streaking_pure_field}
    S_Q(\vb{p}, \tau) = \abs{\bra{\vb{p}} U_\infty(\tau) \ket{\psi(t_0)} \ket{\Phi(t_0)}}^2,
\end{align}
where we defined the long-time limit of the evolution operator $U_\infty(\tau) \equiv \lim_{t \to \infty} U(t;\tau)$. Here, we can already see the significant difference that appears when considering the quantum optical streaking process, namely, instead of computing $\lvert\bra{\vb{p}} \ket{\psi(t)}\rvert^2$ for the the electron wavefunction only, we have in Eq.~\eqref{eq:streaking_pure_field} the additional possibility of interference due to the properties of the initial quantum optical state of the field. 
However, the most interesting insights are obtained when explicitly writing the field state in terms of its phase-space distribution~\cite{schleich2015quantum}, which furthermore allows to explicitly solve the expression of the streaking trace in Eq.~\eqref{eq:streaking_trace} for arbitrary quantum states of light. 
To this end, we express the initial field state in terms of its generalized $P$-representation~\cite{drummond1980generalised}, $P(\alpha, \beta^*)$, such that 
\begin{align}
    \rho_F(t_0) = \int d^2 \alpha \int d^2 \beta \frac{P(\alpha, \beta^*)}{\bra{\beta^*} \ket{\alpha}} \dyad{\alpha}{\beta^*}.
\end{align}

With this we can write the exact quantum optical streaking trace as
\begin{align}
\label{eq:streaking_final}
    S_Q(\vb{p}, \tau) = \int d^2 \alpha \int d^2 \beta \, P(\alpha, \beta^*) M_\alpha(\vb{p}, \tau) M_{\beta^*}^*(\vb{p}, \tau),
\end{align}
where $M_\alpha(\vb{p}, \tau)$ are the semi-classical streaking amplitudes. 
This is the exact expression of the streaking trace for an arbitrary IR streaking field, where the quantum nature and its fluctuations are manifested through the initial IR phase-space distribution $P_{IR}(\alpha, \beta^*)$, and the interference of the streaking amplitudes $M_\alpha(\vb{p}, \tau)$. 
The interference of the different contributions allows to measure the quantum fluctuations of light on the sub-femtosecond time-scale.
The striking difference between the semi-classical streaking trace for a classical coherent field $S_\alpha (\vb{p},\tau) = \abs{M_\alpha(\vb{p}, \tau)}^2$ and the quantum optical version $S_Q(\vb{p}, \tau)$ in Eq.~\eqref{eq:streaking_final}, is the presence of interferences between different classical streaking channels in the latter, corresponding to streaking amplitudes with fields $\alpha \neq \beta^*$. 
The phase-space distribution of the IR quantum state, $P_{IR}(\alpha, \beta^*)$, is given by the generalized P-representation~\cite{drummond1980generalised}, which has the advantage to be a unique, positive and finite distribution function~\cite{gilchrist1997positive, kim1989properties, olsen2009numerical}.
Now, to get a better understanding of the physical process we take into account the large photon numbers of the IR field, and consider the quasi-classical limit of the positive $P$ function~\cite{gorlach2023high, stammer2025weak, rivera2026attosecond, rivera2025structured, gothelf2026limitations}. With this, we can write the quantum optical streaking trace as 
\begin{align}
\label{eq:streaking_Q}
    S_Q(\vb{p}, \tau) = \int d^2 \alpha Q(\alpha) \abs{M_\alpha(\vb{p},\tau)}^2,
\end{align}
where $Q(\alpha) = \frac{1}{\pi} \bra{\alpha} \rho_{IR}\ket{\alpha}$ is the Husimi distribution of the IR state. Although the quasi-classical limit does not lead to interference effects between different contributions of $\alpha \neq \beta^*$, the quantum noise is still taken into account via $Q(\alpha)$. Furthermore, it was shown that for strong field processes the difference between the averages via $P(\alpha, \beta^*)$ and $Q(\alpha)$ do not differ due to the exponential decay of the coherent state overlaps~\cite{wang2025high}.

This formalism enables us not only to perform the measurement of quantum noise on the attosecond time-scale, but can further be seen as a new approach for state tomography of bright quantum light, resolving the long-standing certification problem of intense squeezed light sources~\cite{raymer20047}. In the following, we discuss the features of this new scheme and its implications on the ability to measure the quantum fluctuations of light.

\begin{figure}
    \centering
	\includegraphics[width= 1.0 \columnwidth]{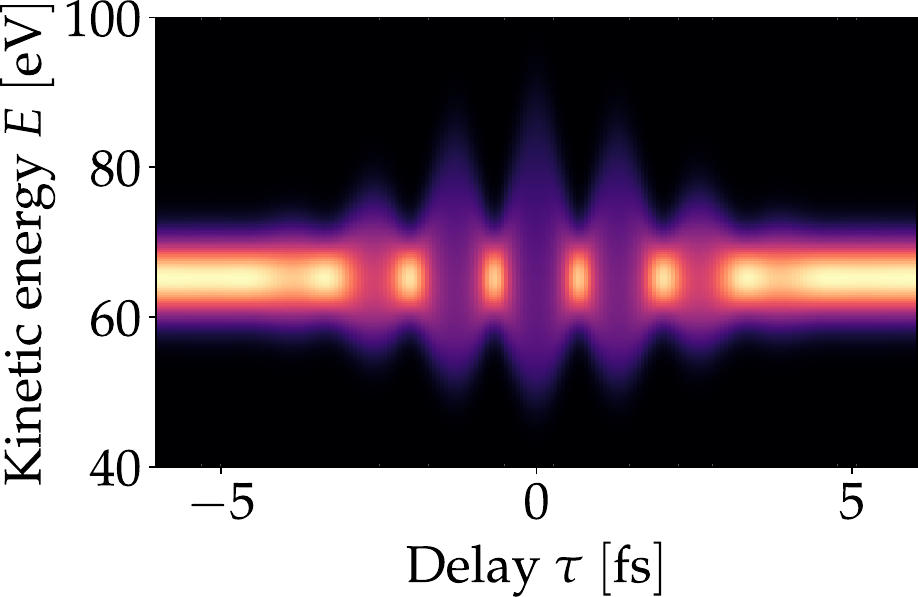}
	\caption{\textbf{Attosecond streaking a bright squeezed vacuum field.} Quantum optical streaking trace of a BSV field showing distinct sub-cycle $2\omega$ oscillations around a vanishing mean value of the streaking field with prominent quantum fluctuations.~Here, we set the squeezing intensity of the streaking field ($\omega = 0.057$ a.u.)~to $I_{\text{squ}} = 5 \times 10^{-5}$ a.u., using helium ($I_p = 0.904$ a.u.) for the atomic system, and used a coherent XUV pulse of $\Omega = 3.3$ a.u.~and IR pulse duration (FWHM) of 5 fs. }
      \label{fig:BSV_streaking}
\end{figure}

\textit{Quantum optical attosecond streaking trace.--}
The general expression of Eq.~\eqref{eq:streaking_final}, allows to compute the QOST for arbitrary IR fields. Hence, we show in Fig.~\ref{fig:BSV_streaking} the QOST for a bright squeezed vacuum (BSV) field $\ket{\xi}$, widely used experimentally to drive strong field processes with quantum light~\cite{rasputnyi2024high, heimerl_multiphoton_2024, heimerl2025quantum}.
We can see that the squeezing in the IR field significantly modifies the streaking trace via distinct $2\omega$ oscillations, and delay-depend field fluctuations. The increased fluctuations occur when the XUV burst probes the IR streaking field at instances $\tau$ when the field is oriented along the anti-squeezed quadrature, while the minima in the width of the streaking trace probe the field along the quantum squeezed quadrature. 
Interestingly, the mean oscillations vanish for the BSV streaking field~\cite{stammer2024limitations}, due to the absence of a classical average electric field $\mathbb{E}_\alpha [\vb{E}_\alpha(t)] = 0$, where the ensemble $\mathbb{E}_\alpha[\cdot ]$ is taken over the distribution of coherent states in the initial IR field. 

\begin{figure}
    \centering
	\includegraphics[width= 1.0 \columnwidth]{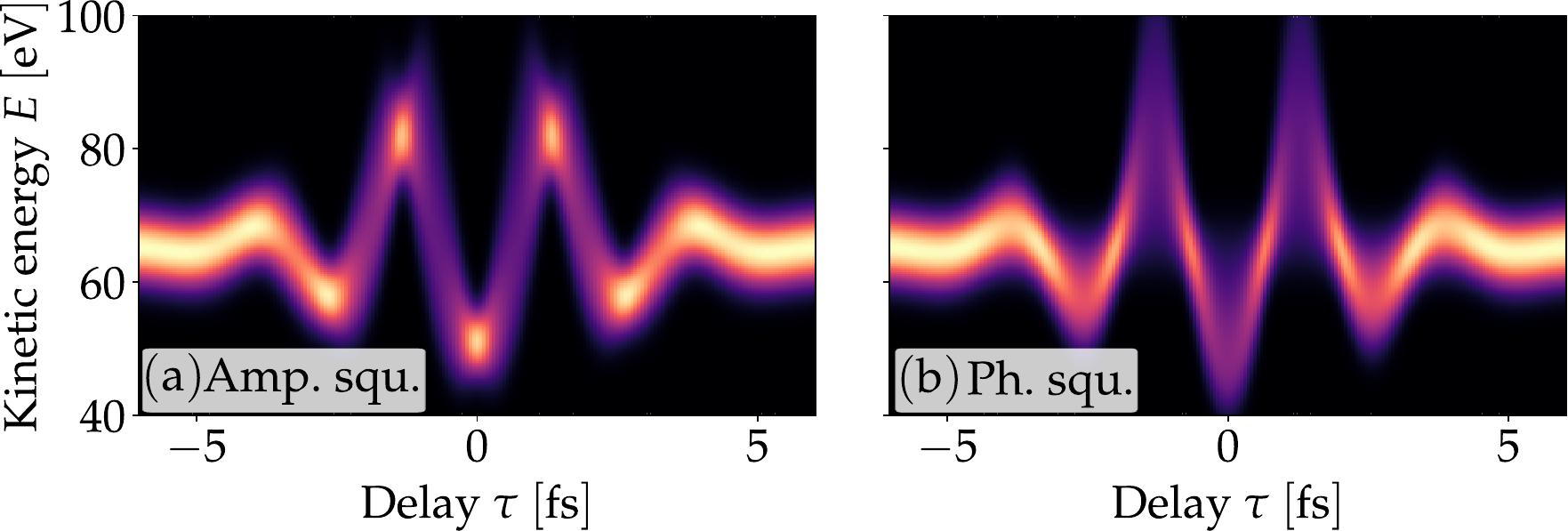}
	\caption{\textbf{Attosecond streaking trace of bright quantum light.} Quantum optical streaking trace for a displaced squeezed state $D(\alpha)\ket{\xi}$ with amplitude squeezing in (a) and phase squeezing in (b). The squeezing intensity of $I_{sq} = 5 \times 10^{-5}$ a.u. is the same for all cases. For the amplitude and phase squeezed state we considered the ratio of $I_{coh}/I_{sq} \sim 6$, for the intensity from the coherent and the squeezed contribution.}
      \label{fig:streaking_quantum_light}
\end{figure}

To further understand the implications on the attosecond streaking measurement for different IR quantum fields, we show in Fig.~\ref{fig:streaking_quantum_light}~(a) and (b) the streaking trace of an amplitude and phase squeezed state, respectively.
Both fields have a definite phase-reference due to the displacement in phase-space, leading to coherent mean field oscillations such that $\mathbb{E}_\alpha [\vb{E}_\alpha(t)] \neq 0$, and follow the mean value of the classical streaking field with frequency $\omega$.
However, both fields have significantly different modulations in their variance due to the fluctuations being distributed differently along the field quadratures. While the amplitude squeezed field has its minimal fluctuations at the intensity maxima, the phase squeezed field has reduced quadrature fluctuations at the zeros of the mean field oscillations. These characteristics of the quadrature fluctuations are translated to the variance of the quantum optical streaking trace, directly showing the ability to witness the quantum fluctuations. 

In summary, although the streaking trace in Eq.~\eqref{eq:streaking_Q} samples over the quantum noise distribution $Q(\alpha)$, the fluctuations of the IR streaking fields are still present in the shot-to-shot data providing a direct measurement of the underlying quantum fluctuations. 
In the following the show how the quantum optical streaking trace allows to measure quantum fluctuations of light.

\textit{Measuring the quantum noise of light.--}
All approaches to detect bright quantum light are thus far limited to the measurement of the $g^{(2)}(0)$ function, where values above that of thermal fluctuations $g^{(2)}(0) > 2$, indicate a super-Poissonian photon number distribution with its long tail typical for squeezed light. In fact, for a perfect and pure BSV field the second-order correlation function reaches $g^{(2)}(0)=3$. However, there also exist classical fields having the same $g^{(2)}$ function, such that the broad photon number distribution and large $g^{(2)}$ values are no unambiguous squeezing signature. For instance, the broad photon number distribution also appears for classical fields with anti-squeezing in one quadrature while the conjugate quadrate has fluctuations above the vacuum limit, such that the $g^{(2)}(0)$ measurement alone is no direct evidence for quantum light.
In consequence, the existing measurements~\cite{lemieux2024photon, rasputnyi2024high, heimerl_multiphoton_2024, heimerl2025quantum, kern2026single}, although indicating super-Poissonian statistics, are inconclusive about quantum squeezing and do not serve as a full quantum state reconstruction method.
In the following, we show that a squeezed field has distinct signatures compared to classical light fields.

\begin{figure}
    \centering
	\includegraphics[width=0.9\columnwidth]{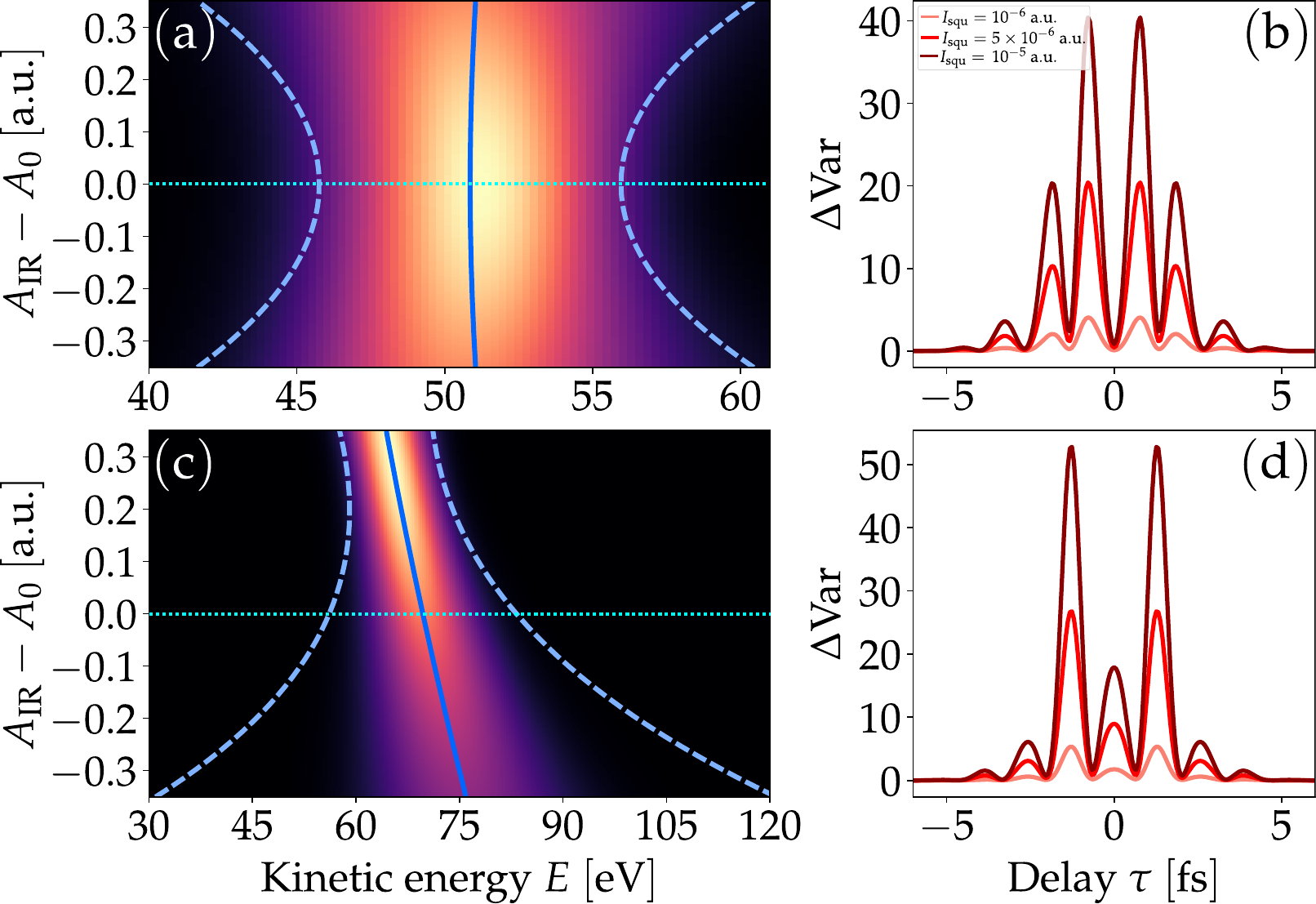}
	\caption{\textbf{Physical origin of the variance distribution for anti-squeezed light.} Streaking trace at fixed time-delay $\tau$ for the classical fields resembling the amplitude and phase-squeezed vector potential. For (a) and (b) the field $A_{\text{am}}(t)$, while for (c) and (d) the field $A_{\text{ph}}(t)$ was considered.}
      \label{fig:anti-squeezing}
\end{figure}

Therefore, we are now interested to understand the contribution of the field fluctuations in the field quadratures.
Thus, we consider a model that incorporates the anti-squeezed quadrature in a classical streaking field. For amplitude squeezing the IR vector potential is considered to be 
\begin{align}
    A_{\text{am}}(t) = f(t) \left[ A_0 \cos(\omega t) + A_{IR} \cos(\omega t) \right],
\end{align}
which effectively introduces an increase in the CEP phase, like the increasing variance in the conjugate phase quadrature. For the phase squeezed streaking field we used 
\begin{align}
    A_{\text{ph}}(t) = f(t) \left( A_0 + A_{IR} \right) \cos(\omega t),
\end{align}
leading to an increase in field amplitude while keeping the phase is fixed. 
Now, we consider these two classical fields, that incorporate the anti-squeezed quadrature of the field, and compute the streaking trace. In particular, we evaluate the streaking trace and its variance at a fixed time-delay as shown in Fig.~\ref{fig:anti-squeezing}. 
For the amplitude squeezed field $A_{\text{am}}(t)$ this is done at $\tau=0$, where the symmetry in Fig.~\ref{fig:anti-squeezing}~(a) reflects the symmetry of the sampling of $A_{IR}$. For phase squeezing $A_{\text{ph}}(t)$ the streaking trace is sampled at $\tau \approx 0.78$ fs, when the vector potential is zero. In this case, a change in $A_{IR}$ keeps the phase constant wile the amplitudes shifts to the positive or negative direction leading to the asymmetry in Fig.~\ref{fig:anti-squeezing}~(c).
However, what we can see is that these fields lead to distinct streaking traces compared to a classical field $A_{\text{cl}} = f(t) A_0 \cos(\omega t)$, with non-trivial modulations in their distribution.
This can be seen in Fig.~\ref{fig:anti-squeezing}~(b) and (d), where we show the difference of the streaking trace variance $\Delta \text{Var} = \text{Var}_{\text{am/ph}} - \text{Var}_{\text{cl}}$, between the amplitude ($A_{\text{am}}$) or phase squeezed ($A_{\text{ph}}$) field with that of the classical field $A_{\text{cl}}$. We can directly see that the presence of the anti-squeezed fluctuations leads to a larger variance in the streaking trace compared to that of a classical coherent field.

\textit{Conclusion.--}
In this work we open the path towards new certification methods of bright quantum light using attosecond metrology techniques. We have demonstrated that the measurement of the quantum optical streaking trace is sensitive to the fluctuations of bright quantum light. The main prediction of this work is that the streaking trace allows to measure quantum squeezing with distinct features from classical light.  
The proposed scheme is already compatible with current state of the art experiments using bright squeezed vacuum sources, providing an alternative approach towards quantum tomography in otherwise inaccessible regimes. 
Here, it is interesting to note that the time-delay $\tau$ between the XUV pulse and IR streaking field share a similar role than the relative phase $\phi$ between the signal field and the local oscillator in a homodyne measurement configuration~\cite{thesis_stammer}. This particular similarity allows the streaking measurement to observe the sub-cycle attosecond quantum fluctuations of light.

\begin{acknowledgments}

P.S. gratefully acknowledges stimulating discussions with Anne L'Huillier.
ICFO group acknowledges support from: Ministerio de Ciencia y Innovation Agencia Estatal de Investigaciones (R$\&$D project CEX2019-000910-S, AEI/10.13039/501100011033, Plan National FIDEUA PID2019-106901GB-I00, FPI), Fundació Privada Cellex, Fundació Mir-Puig, and from Generalitat de Catalunya (AGAUR Grant No. 2017 SGR 1341, CERCA program), and MICIIN with funding from European Union NextGenerationEU(PRTR-C17.I1) and by Generalitat de Catalunya and EU Horizon 2020 FET-OPEN OPTOlogic (Grant No 899794) and ERC AdG NOQIA.
J.R.-D. acknowledges funding from UK Engineering and Physical Sciences Research Council (EPSRC) Funding, Grant UKRI2300 - Attosecond Photoelectron Imaging with Quantum Light.
E.P. acknowledges Royal Society funding under URF\textbackslash R1\textbackslash 211390.

\end{acknowledgments}

\bibliography{literatur}{}

\begin{thebibliography}{78}%
\makeatletter
\providecommand \@ifxundefined [1]{%
 \@ifx{#1\undefined}
}%
\providecommand \@ifnum [1]{%
 \ifnum #1\expandafter \@firstoftwo
 \else \expandafter \@secondoftwo
 \fi
}%
\providecommand \@ifx [1]{%
 \ifx #1\expandafter \@firstoftwo
 \else \expandafter \@secondoftwo
 \fi
}%
\providecommand \natexlab [1]{#1}%
\providecommand \enquote  [1]{``#1''}%
\providecommand \bibnamefont  [1]{#1}%
\providecommand \bibfnamefont [1]{#1}%
\providecommand \citenamefont [1]{#1}%
\providecommand \href@noop [0]{\@secondoftwo}%
\providecommand \href [0]{\begingroup \@sanitize@url \@href}%
\providecommand \@href[1]{\@@startlink{#1}\@@href}%
\providecommand \@@href[1]{\endgroup#1\@@endlink}%
\providecommand \@sanitize@url [0]{\catcode `\\12\catcode `\$12\catcode `\&12\catcode `\#12\catcode `\^12\catcode `\_12\catcode `\%12\relax}%
\providecommand \@@startlink[1]{}%
\providecommand \@@endlink[0]{}%
\providecommand \url  [0]{\begingroup\@sanitize@url \@url }%
\providecommand \@url [1]{\endgroup\@href {#1}{\urlprefix }}%
\providecommand \urlprefix  [0]{URL }%
\providecommand \Eprint [0]{\href }%
\providecommand \doibase [0]{https://doi.org/}%
\providecommand \selectlanguage [0]{\@gobble}%
\providecommand \bibinfo  [0]{\@secondoftwo}%
\providecommand \bibfield  [0]{\@secondoftwo}%
\providecommand \translation [1]{[#1]}%
\providecommand \BibitemOpen [0]{}%
\providecommand \bibitemStop [0]{}%
\providecommand \bibitemNoStop [0]{.\EOS\space}%
\providecommand \EOS [0]{\spacefactor3000\relax}%
\providecommand \BibitemShut  [1]{\csname bibitem#1\endcsname}%
\let\auto@bib@innerbib\@empty
\bibitem [{\citenamefont {Hentschel}\ \emph {et~al.}(2001)\citenamefont {Hentschel}, \citenamefont {Kienberger}, \citenamefont {Spielmann}, \citenamefont {Reider}, \citenamefont {Milosevic}, \citenamefont {Brabec}, \citenamefont {Corkum}, \citenamefont {Heinzmann}, \citenamefont {Drescher},\ and\ \citenamefont {Krausz}}]{hentschel2001attosecond}%
  \BibitemOpen
  \bibfield  {author} {\bibinfo {author} {\bibfnamefont {M.}~\bibnamefont {Hentschel}}, \bibinfo {author} {\bibfnamefont {R.}~\bibnamefont {Kienberger}}, \bibinfo {author} {\bibfnamefont {C.}~\bibnamefont {Spielmann}}, \bibinfo {author} {\bibfnamefont {G.~A.}\ \bibnamefont {Reider}}, \bibinfo {author} {\bibfnamefont {N.}~\bibnamefont {Milosevic}}, \bibinfo {author} {\bibfnamefont {T.}~\bibnamefont {Brabec}}, \bibinfo {author} {\bibfnamefont {P.}~\bibnamefont {Corkum}}, \bibinfo {author} {\bibfnamefont {U.}~\bibnamefont {Heinzmann}}, \bibinfo {author} {\bibfnamefont {M.}~\bibnamefont {Drescher}},\ and\ \bibinfo {author} {\bibfnamefont {F.}~\bibnamefont {Krausz}},\ }\bibfield  {title} {\bibinfo {title} {Attosecond metrology},\ }\href {https://www.nature.com/articles/35107000} {\bibfield  {journal} {\bibinfo  {journal} {Nature}\ }\textbf {\bibinfo {volume} {414}},\ \bibinfo {pages} {509} (\bibinfo {year} {2001})}\BibitemShut {NoStop}%
\bibitem [{\citenamefont {Drescher}\ \emph {et~al.}(2001)\citenamefont {Drescher}, \citenamefont {Hentschel}, \citenamefont {Kienberger}, \citenamefont {Tempea}, \citenamefont {Spielmann}, \citenamefont {Reider}, \citenamefont {Corkum},\ and\ \citenamefont {Krausz}}]{drescher2001x}%
  \BibitemOpen
  \bibfield  {author} {\bibinfo {author} {\bibfnamefont {M.}~\bibnamefont {Drescher}}, \bibinfo {author} {\bibfnamefont {M.}~\bibnamefont {Hentschel}}, \bibinfo {author} {\bibfnamefont {R.}~\bibnamefont {Kienberger}}, \bibinfo {author} {\bibfnamefont {G.}~\bibnamefont {Tempea}}, \bibinfo {author} {\bibfnamefont {C.}~\bibnamefont {Spielmann}}, \bibinfo {author} {\bibfnamefont {G.~A.}\ \bibnamefont {Reider}}, \bibinfo {author} {\bibfnamefont {P.~B.}\ \bibnamefont {Corkum}},\ and\ \bibinfo {author} {\bibfnamefont {F.}~\bibnamefont {Krausz}},\ }\bibfield  {title} {\bibinfo {title} {X-ray pulses approaching the attosecond frontier},\ }\href {https://www.science.org/doi/10.1126/science.1058561} {\bibfield  {journal} {\bibinfo  {journal} {Science}\ }\textbf {\bibinfo {volume} {291}},\ \bibinfo {pages} {1923} (\bibinfo {year} {2001})}\BibitemShut {NoStop}%
\bibitem [{\citenamefont {Kienberger}\ \emph {et~al.}(2004)\citenamefont {Kienberger}, \citenamefont {Goulielmakis}, \citenamefont {Uiberacker}, \citenamefont {Baltuska}, \citenamefont {Yakovlev}, \citenamefont {Bammer}, \citenamefont {Scrinzi}, \citenamefont {Westerwalbesloh}, \citenamefont {Kleineberg}, \citenamefont {Heinzmann} \emph {et~al.}}]{kienberger2004atomic}%
  \BibitemOpen
  \bibfield  {author} {\bibinfo {author} {\bibfnamefont {R.}~\bibnamefont {Kienberger}}, \bibinfo {author} {\bibfnamefont {E.}~\bibnamefont {Goulielmakis}}, \bibinfo {author} {\bibfnamefont {M.}~\bibnamefont {Uiberacker}}, \bibinfo {author} {\bibfnamefont {A.}~\bibnamefont {Baltuska}}, \bibinfo {author} {\bibfnamefont {V.}~\bibnamefont {Yakovlev}}, \bibinfo {author} {\bibfnamefont {F.}~\bibnamefont {Bammer}}, \bibinfo {author} {\bibfnamefont {A.}~\bibnamefont {Scrinzi}}, \bibinfo {author} {\bibfnamefont {T.}~\bibnamefont {Westerwalbesloh}}, \bibinfo {author} {\bibfnamefont {U.}~\bibnamefont {Kleineberg}}, \bibinfo {author} {\bibfnamefont {U.}~\bibnamefont {Heinzmann}}, \emph {et~al.},\ }\bibfield  {title} {\bibinfo {title} {Atomic transient recorder},\ }\href {https://www.nature.com/articles/nature02277} {\bibfield  {journal} {\bibinfo  {journal} {Nature}\ }\textbf {\bibinfo {volume} {427}},\ \bibinfo {pages} {817} (\bibinfo {year} {2004})}\BibitemShut {NoStop}%
\bibitem [{\citenamefont {Krausz}\ and\ \citenamefont {Ivanov}(2009)}]{krausz2009attosecond}%
  \BibitemOpen
  \bibfield  {author} {\bibinfo {author} {\bibfnamefont {F.}~\bibnamefont {Krausz}}\ and\ \bibinfo {author} {\bibfnamefont {M.}~\bibnamefont {Ivanov}},\ }\bibfield  {title} {\bibinfo {title} {Attosecond physics},\ }\href {https://journals.aps.org/rmp/abstract/10.1103/RevModPhys.81.163} {\bibfield  {journal} {\bibinfo  {journal} {Reviews of Modern Physics}\ }\textbf {\bibinfo {volume} {81}},\ \bibinfo {pages} {163} (\bibinfo {year} {2009})}\BibitemShut {NoStop}%
\bibitem [{\citenamefont {Mairesse}\ and\ \citenamefont {Qu{\'e}r{\'e}}(2005)}]{mairesse2005frequency}%
  \BibitemOpen
  \bibfield  {author} {\bibinfo {author} {\bibfnamefont {Y.}~\bibnamefont {Mairesse}}\ and\ \bibinfo {author} {\bibfnamefont {F.}~\bibnamefont {Qu{\'e}r{\'e}}},\ }\bibfield  {title} {\bibinfo {title} {Frequency-resolved optical gating for complete reconstruction of attosecond bursts},\ }\href {https://journals.aps.org/pra/abstract/10.1103/PhysRevA.71.011401} {\bibfield  {journal} {\bibinfo  {journal} {Physical Review A—Atomic, Molecular, and Optical Physics}\ }\textbf {\bibinfo {volume} {71}},\ \bibinfo {pages} {011401} (\bibinfo {year} {2005})}\BibitemShut {NoStop}%
\bibitem [{\citenamefont {Qu{\'e}r{\'e}}\ \emph {et~al.}(2005)\citenamefont {Qu{\'e}r{\'e}}, \citenamefont {Mairesse},\ and\ \citenamefont {Itatani}}]{quere2005temporal}%
  \BibitemOpen
  \bibfield  {author} {\bibinfo {author} {\bibfnamefont {F.}~\bibnamefont {Qu{\'e}r{\'e}}}, \bibinfo {author} {\bibfnamefont {Y.}~\bibnamefont {Mairesse}},\ and\ \bibinfo {author} {\bibfnamefont {J.}~\bibnamefont {Itatani}},\ }\bibfield  {title} {\bibinfo {title} {Temporal characterization of attosecond xuv fields},\ }\href {https://www.tandfonline.com/doi/full/10.1080/09500340412331307942} {\bibfield  {journal} {\bibinfo  {journal} {Journal of Modern Optics}\ }\textbf {\bibinfo {volume} {52}},\ \bibinfo {pages} {339} (\bibinfo {year} {2005})}\BibitemShut {NoStop}%
\bibitem [{\citenamefont {Schultze}\ \emph {et~al.}(2010)\citenamefont {Schultze}, \citenamefont {Fie{\ss}}, \citenamefont {Karpowicz}, \citenamefont {Gagnon}, \citenamefont {Korbman}, \citenamefont {Hofstetter}, \citenamefont {Neppl}, \citenamefont {Cavalieri}, \citenamefont {Komninos}, \citenamefont {Mercouris} \emph {et~al.}}]{schultze2010delay}%
  \BibitemOpen
  \bibfield  {author} {\bibinfo {author} {\bibfnamefont {M.}~\bibnamefont {Schultze}}, \bibinfo {author} {\bibfnamefont {M.}~\bibnamefont {Fie{\ss}}}, \bibinfo {author} {\bibfnamefont {N.}~\bibnamefont {Karpowicz}}, \bibinfo {author} {\bibfnamefont {J.}~\bibnamefont {Gagnon}}, \bibinfo {author} {\bibfnamefont {M.}~\bibnamefont {Korbman}}, \bibinfo {author} {\bibfnamefont {M.}~\bibnamefont {Hofstetter}}, \bibinfo {author} {\bibfnamefont {S.}~\bibnamefont {Neppl}}, \bibinfo {author} {\bibfnamefont {A.~L.}\ \bibnamefont {Cavalieri}}, \bibinfo {author} {\bibfnamefont {Y.}~\bibnamefont {Komninos}}, \bibinfo {author} {\bibfnamefont {T.}~\bibnamefont {Mercouris}}, \emph {et~al.},\ }\bibfield  {title} {\bibinfo {title} {Delay in photoemission},\ }\href {https://www.science.org/doi/full/10.1126/science.1189401} {\bibfield  {journal} {\bibinfo  {journal} {Science}\ }\textbf {\bibinfo {volume} {328}},\ \bibinfo {pages} {1658} (\bibinfo {year} {2010})}\BibitemShut {NoStop}%
\bibitem [{\citenamefont {Itatani}\ \emph {et~al.}(2002)\citenamefont {Itatani}, \citenamefont {Qu{\'e}r{\'e}}, \citenamefont {Yudin}, \citenamefont {Ivanov}, \citenamefont {Krausz},\ and\ \citenamefont {Corkum}}]{itatani2002attosecond}%
  \BibitemOpen
  \bibfield  {author} {\bibinfo {author} {\bibfnamefont {J.}~\bibnamefont {Itatani}}, \bibinfo {author} {\bibfnamefont {F.}~\bibnamefont {Qu{\'e}r{\'e}}}, \bibinfo {author} {\bibfnamefont {G.~L.}\ \bibnamefont {Yudin}}, \bibinfo {author} {\bibfnamefont {M.~Y.}\ \bibnamefont {Ivanov}}, \bibinfo {author} {\bibfnamefont {F.}~\bibnamefont {Krausz}},\ and\ \bibinfo {author} {\bibfnamefont {P.~B.}\ \bibnamefont {Corkum}},\ }\bibfield  {title} {\bibinfo {title} {Attosecond streak camera},\ }\href {https://journals.aps.org/prl/abstract/10.1103/PhysRevLett.88.173903} {\bibfield  {journal} {\bibinfo  {journal} {Physical review letters}\ }\textbf {\bibinfo {volume} {88}},\ \bibinfo {pages} {173903} (\bibinfo {year} {2002})}\BibitemShut {NoStop}%
\bibitem [{\citenamefont {Kitzler}\ \emph {et~al.}(2002)\citenamefont {Kitzler}, \citenamefont {Milosevic}, \citenamefont {Scrinzi}, \citenamefont {Krausz},\ and\ \citenamefont {Brabec}}]{kitzler2002quantum}%
  \BibitemOpen
  \bibfield  {author} {\bibinfo {author} {\bibfnamefont {M.}~\bibnamefont {Kitzler}}, \bibinfo {author} {\bibfnamefont {N.}~\bibnamefont {Milosevic}}, \bibinfo {author} {\bibfnamefont {A.}~\bibnamefont {Scrinzi}}, \bibinfo {author} {\bibfnamefont {F.}~\bibnamefont {Krausz}},\ and\ \bibinfo {author} {\bibfnamefont {T.}~\bibnamefont {Brabec}},\ }\bibfield  {title} {\bibinfo {title} {Quantum theory of attosecond xuv pulse measurement by laser dressed photoionization},\ }\href {https://journals.aps.org/prl/abstract/10.1103/PhysRevLett.88.173904} {\bibfield  {journal} {\bibinfo  {journal} {Physical review letters}\ }\textbf {\bibinfo {volume} {88}},\ \bibinfo {pages} {173904} (\bibinfo {year} {2002})}\BibitemShut {NoStop}%
\bibitem [{\citenamefont {Goulielmakis}\ \emph {et~al.}(2004)\citenamefont {Goulielmakis}, \citenamefont {Uiberacker}, \citenamefont {Kienberger}, \citenamefont {Baltuska}, \citenamefont {Yakovlev}, \citenamefont {Scrinzi}, \citenamefont {Westerwalbesloh}, \citenamefont {Kleineberg}, \citenamefont {Heinzmann}, \citenamefont {Drescher} \emph {et~al.}}]{goulielmakis2004direct}%
  \BibitemOpen
  \bibfield  {author} {\bibinfo {author} {\bibfnamefont {E.}~\bibnamefont {Goulielmakis}}, \bibinfo {author} {\bibfnamefont {M.}~\bibnamefont {Uiberacker}}, \bibinfo {author} {\bibfnamefont {R.}~\bibnamefont {Kienberger}}, \bibinfo {author} {\bibfnamefont {A.}~\bibnamefont {Baltuska}}, \bibinfo {author} {\bibfnamefont {V.}~\bibnamefont {Yakovlev}}, \bibinfo {author} {\bibfnamefont {A.}~\bibnamefont {Scrinzi}}, \bibinfo {author} {\bibfnamefont {T.}~\bibnamefont {Westerwalbesloh}}, \bibinfo {author} {\bibfnamefont {U.}~\bibnamefont {Kleineberg}}, \bibinfo {author} {\bibfnamefont {U.}~\bibnamefont {Heinzmann}}, \bibinfo {author} {\bibfnamefont {M.}~\bibnamefont {Drescher}}, \emph {et~al.},\ }\bibfield  {title} {\bibinfo {title} {Direct measurement of light waves},\ }\href {https://www.science.org/doi/10.1126/science.1100866} {\bibfield  {journal} {\bibinfo  {journal} {Science}\ }\textbf {\bibinfo {volume} {305}},\ \bibinfo {pages} {1267} (\bibinfo {year} {2004})}\BibitemShut {NoStop}%
\bibitem [{\citenamefont {Corkum}\ and\ \citenamefont {Krausz}(2007)}]{corkum2007attosecond}%
  \BibitemOpen
  \bibfield  {author} {\bibinfo {author} {\bibfnamefont {P.~B.}\ \bibnamefont {Corkum}}\ and\ \bibinfo {author} {\bibfnamefont {F.}~\bibnamefont {Krausz}},\ }\bibfield  {title} {\bibinfo {title} {Attosecond science},\ }\href {https://www.nature.com/articles/nphys620} {\bibfield  {journal} {\bibinfo  {journal} {Nature Physics}\ }\textbf {\bibinfo {volume} {3}},\ \bibinfo {pages} {381} (\bibinfo {year} {2007})}\BibitemShut {NoStop}%
\bibitem [{\citenamefont {L’Huillier}(2024)}]{Huillier_Nobel_2024}%
  \BibitemOpen
  \bibfield  {author} {\bibinfo {author} {\bibfnamefont {A.}~\bibnamefont {L’Huillier}},\ }\bibfield  {title} {\bibinfo {title} {Nobel lecture: The route to attosecond pulses},\ }\href {https://doi.org/10.1103/RevModPhys.96.030503} {\bibfield  {journal} {\bibinfo  {journal} {Rev. Mod. Phys.}\ }\textbf {\bibinfo {volume} {96}},\ \bibinfo {pages} {030503} (\bibinfo {year} {2024})}\BibitemShut {NoStop}%
\bibitem [{\citenamefont {Agostini}(2024)}]{Agostini_Nobel_2024}%
  \BibitemOpen
  \bibfield  {author} {\bibinfo {author} {\bibfnamefont {P.}~\bibnamefont {Agostini}},\ }\bibfield  {title} {\bibinfo {title} {Nobel lecture: Genesis and applications of attosecond pulse trains},\ }\href {https://doi.org/10.1103/RevModPhys.96.030501} {\bibfield  {journal} {\bibinfo  {journal} {Rev. Mod. Phys.}\ }\textbf {\bibinfo {volume} {96}},\ \bibinfo {pages} {030501} (\bibinfo {year} {2024})}\BibitemShut {NoStop}%
\bibitem [{\citenamefont {Krausz}(2024)}]{Krausz_Nobel_2024}%
  \BibitemOpen
  \bibfield  {author} {\bibinfo {author} {\bibfnamefont {F.}~\bibnamefont {Krausz}},\ }\bibfield  {title} {\bibinfo {title} {Nobel lecture: Sub-atomic motions},\ }\href {https://doi.org/10.1103/RevModPhys.96.030502} {\bibfield  {journal} {\bibinfo  {journal} {Rev. Mod. Phys.}\ }\textbf {\bibinfo {volume} {96}},\ \bibinfo {pages} {030502} (\bibinfo {year} {2024})}\BibitemShut {NoStop}%
\bibitem [{\citenamefont {Stammer}(2026)}]{thesis_stammer}%
  \BibitemOpen
  \bibfield  {author} {\bibinfo {author} {\bibfnamefont {P.}~\bibnamefont {Stammer}},\ }\emph {\bibinfo {title} {Photons and Information -- A modern approach to strong-field quantum optics}},\ \href@noop {} {Ph.D. thesis},\ \bibinfo  {school} {ICFO} (\bibinfo {year} {2026})\BibitemShut {NoStop}%
\bibitem [{\citenamefont {Wiseman}\ and\ \citenamefont {Milburn}(1993)}]{wiseman1993quantum}%
  \BibitemOpen
  \bibfield  {author} {\bibinfo {author} {\bibfnamefont {H.~M.}\ \bibnamefont {Wiseman}}\ and\ \bibinfo {author} {\bibfnamefont {G.~J.}\ \bibnamefont {Milburn}},\ }\bibfield  {title} {\bibinfo {title} {Quantum theory of field-quadrature measurements},\ }\href {https://journals.aps.org/pra/abstract/10.1103/PhysRevA.47.642} {\bibfield  {journal} {\bibinfo  {journal} {Physical review A}\ }\textbf {\bibinfo {volume} {47}},\ \bibinfo {pages} {642} (\bibinfo {year} {1993})}\BibitemShut {NoStop}%
\bibitem [{\citenamefont {Breitenbach}\ and\ \citenamefont {Schiller}(1997)}]{breitenbach1997homodyne}%
  \BibitemOpen
  \bibfield  {author} {\bibinfo {author} {\bibfnamefont {G.}~\bibnamefont {Breitenbach}}\ and\ \bibinfo {author} {\bibfnamefont {S.}~\bibnamefont {Schiller}},\ }\bibfield  {title} {\bibinfo {title} {Homodyne tomography of classical and non-classical light},\ }\href {https://www.tandfonline.com/doi/abs/10.1080/09500349708231879?casa_token=ha6TSC0R1tIAAAAA:xw1fWC-YaT3fUGOUe8SbS5vMJPAgE5m4XBFsQ7vroNg5mZ88BY7MpiSpJinm3el-Dv6lA3UopJMetw} {\bibfield  {journal} {\bibinfo  {journal} {Journal of Modern Optics}\ }\textbf {\bibinfo {volume} {44}},\ \bibinfo {pages} {2207} (\bibinfo {year} {1997})}\BibitemShut {NoStop}%
\bibitem [{\citenamefont {Leonhardt}\ and\ \citenamefont {Paul}(1995)}]{leonhardt1995measuring}%
  \BibitemOpen
  \bibfield  {author} {\bibinfo {author} {\bibfnamefont {U.}~\bibnamefont {Leonhardt}}\ and\ \bibinfo {author} {\bibfnamefont {H.}~\bibnamefont {Paul}},\ }\bibfield  {title} {\bibinfo {title} {Measuring the quantum state of light},\ }\href {https://www.sciencedirect.com/science/article/abs/pii/007967279400007L} {\bibfield  {journal} {\bibinfo  {journal} {Progress in Quantum Electronics}\ }\textbf {\bibinfo {volume} {19}},\ \bibinfo {pages} {89} (\bibinfo {year} {1995})}\BibitemShut {NoStop}%
\bibitem [{\citenamefont {Lvovsky}\ and\ \citenamefont {Raymer}(2009)}]{lvovsky2009continuous}%
  \BibitemOpen
  \bibfield  {author} {\bibinfo {author} {\bibfnamefont {A.~I.}\ \bibnamefont {Lvovsky}}\ and\ \bibinfo {author} {\bibfnamefont {M.~G.}\ \bibnamefont {Raymer}},\ }\bibfield  {title} {\bibinfo {title} {Continuous-variable optical quantum-state tomography},\ }\href {https://journals.aps.org/rmp/abstract/10.1103/RevModPhys.81.299} {\bibfield  {journal} {\bibinfo  {journal} {Reviews of Modern Physics}\ }\textbf {\bibinfo {volume} {81}},\ \bibinfo {pages} {299} (\bibinfo {year} {2009})}\BibitemShut {NoStop}%
\bibitem [{\citenamefont {Schleich}(2015)}]{schleich2015quantum}%
  \BibitemOpen
  \bibfield  {author} {\bibinfo {author} {\bibfnamefont {W.~P.}\ \bibnamefont {Schleich}},\ }\href@noop {} {\emph {\bibinfo {title} {Quantum optics in phase space}}}\ (\bibinfo  {publisher} {John Wiley \& Sons},\ \bibinfo {year} {2015})\BibitemShut {NoStop}%
\bibitem [{\citenamefont {Dowling}\ and\ \citenamefont {Seshadreesan}(2015)}]{dowling2015quantum}%
  \BibitemOpen
  \bibfield  {author} {\bibinfo {author} {\bibfnamefont {J.~P.}\ \bibnamefont {Dowling}}\ and\ \bibinfo {author} {\bibfnamefont {K.~P.}\ \bibnamefont {Seshadreesan}},\ }\bibfield  {title} {\bibinfo {title} {Quantum optical technologies for metrology, sensing, and imaging},\ }\href {https://opg.optica.org/jlt/abstract.cfm?uri=jlt-33-12-2359} {\bibfield  {journal} {\bibinfo  {journal} {Journal of Lightwave Technology}\ }\textbf {\bibinfo {volume} {33}},\ \bibinfo {pages} {2359} (\bibinfo {year} {2015})}\BibitemShut {NoStop}%
\bibitem [{\citenamefont {Dowling}\ and\ \citenamefont {Milburn}(2003)}]{dowling2003quantum}%
  \BibitemOpen
  \bibfield  {author} {\bibinfo {author} {\bibfnamefont {J.~P.}\ \bibnamefont {Dowling}}\ and\ \bibinfo {author} {\bibfnamefont {G.~J.}\ \bibnamefont {Milburn}},\ }\bibfield  {title} {\bibinfo {title} {Quantum technology: the second quantum revolution},\ }\href {https://royalsocietypublishing.org/doi/abs/10.1098/rsta.2003.1227} {\bibfield  {journal} {\bibinfo  {journal} {Philos. Trans. A Math. Phys. Eng. Sci.}\ }\textbf {\bibinfo {volume} {361}},\ \bibinfo {pages} {1655} (\bibinfo {year} {2003})}\BibitemShut {NoStop}%
\bibitem [{\citenamefont {Caves}(1981)}]{caves1981quantum}%
  \BibitemOpen
  \bibfield  {author} {\bibinfo {author} {\bibfnamefont {C.~M.}\ \bibnamefont {Caves}},\ }\bibfield  {title} {\bibinfo {title} {Quantum-mechanical noise in an interferometer},\ }\href {https://journals.aps.org/prd/abstract/10.1103/PhysRevD.23.1693} {\bibfield  {journal} {\bibinfo  {journal} {Physical Review D}\ }\textbf {\bibinfo {volume} {23}},\ \bibinfo {pages} {1693} (\bibinfo {year} {1981})}\BibitemShut {NoStop}%
\bibitem [{\citenamefont {Schnabel}\ \emph {et~al.}(2010)\citenamefont {Schnabel}, \citenamefont {Mavalvala}, \citenamefont {McClelland},\ and\ \citenamefont {Lam}}]{schnabel2010quantum}%
  \BibitemOpen
  \bibfield  {author} {\bibinfo {author} {\bibfnamefont {R.}~\bibnamefont {Schnabel}}, \bibinfo {author} {\bibfnamefont {N.}~\bibnamefont {Mavalvala}}, \bibinfo {author} {\bibfnamefont {D.~E.}\ \bibnamefont {McClelland}},\ and\ \bibinfo {author} {\bibfnamefont {P.~K.}\ \bibnamefont {Lam}},\ }\bibfield  {title} {\bibinfo {title} {Quantum metrology for gravitational wave astronomy},\ }\href {https://www.nature.com/articles/ncomms1122} {\bibfield  {journal} {\bibinfo  {journal} {Nature Communications}\ }\textbf {\bibinfo {volume} {1}},\ \bibinfo {pages} {121} (\bibinfo {year} {2010})}\BibitemShut {NoStop}%
\bibitem [{\citenamefont {Lawrie}\ \emph {et~al.}(2019)\citenamefont {Lawrie}, \citenamefont {Lett}, \citenamefont {Marino},\ and\ \citenamefont {Pooser}}]{lawrie2019quantum}%
  \BibitemOpen
  \bibfield  {author} {\bibinfo {author} {\bibfnamefont {B.~J.}\ \bibnamefont {Lawrie}}, \bibinfo {author} {\bibfnamefont {P.~D.}\ \bibnamefont {Lett}}, \bibinfo {author} {\bibfnamefont {A.~M.}\ \bibnamefont {Marino}},\ and\ \bibinfo {author} {\bibfnamefont {R.~C.}\ \bibnamefont {Pooser}},\ }\bibfield  {title} {\bibinfo {title} {Quantum sensing with squeezed light},\ }\href {https://pubs.acs.org/doi/full/10.1021/acsphotonics.9b00250} {\bibfield  {journal} {\bibinfo  {journal} {Acs Photonics}\ }\textbf {\bibinfo {volume} {6}},\ \bibinfo {pages} {1307} (\bibinfo {year} {2019})}\BibitemShut {NoStop}%
\bibitem [{\citenamefont {Weedbrook}\ \emph {et~al.}(2012)\citenamefont {Weedbrook}, \citenamefont {Pirandola}, \citenamefont {Garc{\'\i}a-Patr{\'o}n}, \citenamefont {Cerf}, \citenamefont {Ralph}, \citenamefont {Shapiro},\ and\ \citenamefont {Lloyd}}]{weedbrook2012gaussian}%
  \BibitemOpen
  \bibfield  {author} {\bibinfo {author} {\bibfnamefont {C.}~\bibnamefont {Weedbrook}}, \bibinfo {author} {\bibfnamefont {S.}~\bibnamefont {Pirandola}}, \bibinfo {author} {\bibfnamefont {R.}~\bibnamefont {Garc{\'\i}a-Patr{\'o}n}}, \bibinfo {author} {\bibfnamefont {N.~J.}\ \bibnamefont {Cerf}}, \bibinfo {author} {\bibfnamefont {T.~C.}\ \bibnamefont {Ralph}}, \bibinfo {author} {\bibfnamefont {J.~H.}\ \bibnamefont {Shapiro}},\ and\ \bibinfo {author} {\bibfnamefont {S.}~\bibnamefont {Lloyd}},\ }\bibfield  {title} {\bibinfo {title} {Gaussian quantum information},\ }\href {https://journals.aps.org/rmp/abstract/10.1103/RevModPhys.84.621} {\bibfield  {journal} {\bibinfo  {journal} {Reviews of Modern Physics}\ }\textbf {\bibinfo {volume} {84}},\ \bibinfo {pages} {621} (\bibinfo {year} {2012})}\BibitemShut {NoStop}%
\bibitem [{\citenamefont {Braunstein}\ and\ \citenamefont {Van~Loock}(2005)}]{braunstein2005quantum}%
  \BibitemOpen
  \bibfield  {author} {\bibinfo {author} {\bibfnamefont {S.~L.}\ \bibnamefont {Braunstein}}\ and\ \bibinfo {author} {\bibfnamefont {P.}~\bibnamefont {Van~Loock}},\ }\bibfield  {title} {\bibinfo {title} {Quantum information with continuous variables},\ }\href {https://journals.aps.org/rmp/abstract/10.1103/RevModPhys.77.513} {\bibfield  {journal} {\bibinfo  {journal} {Reviews of Modern Physics}\ }\textbf {\bibinfo {volume} {77}},\ \bibinfo {pages} {513} (\bibinfo {year} {2005})}\BibitemShut {NoStop}%
\bibitem [{\citenamefont {Usenko}\ \emph {et~al.}(2026)\citenamefont {Usenko}, \citenamefont {Ac{\'\i}n}, \citenamefont {All{\'e}aume}, \citenamefont {Andersen}, \citenamefont {Diamanti}, \citenamefont {Gehring}, \citenamefont {Hajomer}, \citenamefont {Kanitschar}, \citenamefont {Pacher}, \citenamefont {Pirandola} \emph {et~al.}}]{usenko2026continuous}%
  \BibitemOpen
  \bibfield  {author} {\bibinfo {author} {\bibfnamefont {V.~C.}\ \bibnamefont {Usenko}}, \bibinfo {author} {\bibfnamefont {A.}~\bibnamefont {Ac{\'\i}n}}, \bibinfo {author} {\bibfnamefont {R.}~\bibnamefont {All{\'e}aume}}, \bibinfo {author} {\bibfnamefont {U.~L.}\ \bibnamefont {Andersen}}, \bibinfo {author} {\bibfnamefont {E.}~\bibnamefont {Diamanti}}, \bibinfo {author} {\bibfnamefont {T.}~\bibnamefont {Gehring}}, \bibinfo {author} {\bibfnamefont {A.~A.}\ \bibnamefont {Hajomer}}, \bibinfo {author} {\bibfnamefont {F.}~\bibnamefont {Kanitschar}}, \bibinfo {author} {\bibfnamefont {C.}~\bibnamefont {Pacher}}, \bibinfo {author} {\bibfnamefont {S.}~\bibnamefont {Pirandola}}, \emph {et~al.},\ }\bibfield  {title} {\bibinfo {title} {Continuous-variable quantum communication},\ }\href {https://journals.aps.org/rmp/abstract/10.1103/mgj7-t6d3} {\bibfield  {journal} {\bibinfo  {journal} {Reviews of Modern Physics}\ }\textbf {\bibinfo {volume} {98}},\ \bibinfo {pages} {015003} (\bibinfo {year} {2026})}\BibitemShut {NoStop}%
\bibitem [{\citenamefont {Walls}(1983)}]{Walls1983}%
  \BibitemOpen
  \bibfield  {author} {\bibinfo {author} {\bibfnamefont {D.~F.}\ \bibnamefont {Walls}},\ }\bibfield  {title} {\bibinfo {title} {Squeezed states of light},\ }\href {https://doi.org/10.1038/306141a0} {\bibfield  {journal} {\bibinfo  {journal} {Nature}\ }\textbf {\bibinfo {volume} {306}},\ \bibinfo {pages} {141} (\bibinfo {year} {1983})}\BibitemShut {NoStop}%
\bibitem [{\citenamefont {Breitenbach}\ \emph {et~al.}(1997)\citenamefont {Breitenbach}, \citenamefont {Schiller},\ and\ \citenamefont {Mlynek}}]{breitenbach1997measurement}%
  \BibitemOpen
  \bibfield  {author} {\bibinfo {author} {\bibfnamefont {G.}~\bibnamefont {Breitenbach}}, \bibinfo {author} {\bibfnamefont {S.}~\bibnamefont {Schiller}},\ and\ \bibinfo {author} {\bibfnamefont {J.}~\bibnamefont {Mlynek}},\ }\bibfield  {title} {\bibinfo {title} {Measurement of the quantum states of squeezed light},\ }\href {https://www.nature.com/articles/387471a0} {\bibfield  {journal} {\bibinfo  {journal} {Nature}\ }\textbf {\bibinfo {volume} {387}},\ \bibinfo {pages} {471} (\bibinfo {year} {1997})}\BibitemShut {NoStop}%
\bibitem [{\citenamefont {Raymer}\ and\ \citenamefont {Beck}(2004)}]{raymer20047}%
  \BibitemOpen
  \bibfield  {author} {\bibinfo {author} {\bibfnamefont {M.~G.}\ \bibnamefont {Raymer}}\ and\ \bibinfo {author} {\bibfnamefont {M.}~\bibnamefont {Beck}},\ }\bibfield  {title} {\bibinfo {title} {7 experimental quantum state tomography of optical fields and ultrafast statistical sampling},\ }in\ \href {https://link.springer.com/chapter/10.1007/978-3-540-44481-7_7} {\emph {\bibinfo {booktitle} {Quantum State Estimation}}}\ (\bibinfo  {publisher} {Springer},\ \bibinfo {year} {2004})\ pp.\ \bibinfo {pages} {235--295}\BibitemShut {NoStop}%
\bibitem [{\citenamefont {Tiedau}\ \emph {et~al.}(2019)\citenamefont {Tiedau}, \citenamefont {Meyer-Scott}, \citenamefont {Nitsche}, \citenamefont {Barkhofen}, \citenamefont {Bartley},\ and\ \citenamefont {Silberhorn}}]{tiedau2019high}%
  \BibitemOpen
  \bibfield  {author} {\bibinfo {author} {\bibfnamefont {J.}~\bibnamefont {Tiedau}}, \bibinfo {author} {\bibfnamefont {E.}~\bibnamefont {Meyer-Scott}}, \bibinfo {author} {\bibfnamefont {T.}~\bibnamefont {Nitsche}}, \bibinfo {author} {\bibfnamefont {S.}~\bibnamefont {Barkhofen}}, \bibinfo {author} {\bibfnamefont {T.~J.}\ \bibnamefont {Bartley}},\ and\ \bibinfo {author} {\bibfnamefont {C.}~\bibnamefont {Silberhorn}},\ }\bibfield  {title} {\bibinfo {title} {A high dynamic range optical detector for measuring single photons and bright light},\ }\href {https://opg.optica.org/oe/fulltext.cfm?uri=oe-27-1-1} {\bibfield  {journal} {\bibinfo  {journal} {Optics express}\ }\textbf {\bibinfo {volume} {27}},\ \bibinfo {pages} {1} (\bibinfo {year} {2019})}\BibitemShut {NoStop}%
\bibitem [{\citenamefont {Knyazev}\ \emph {et~al.}(2018)\citenamefont {Knyazev}, \citenamefont {Spasibko}, \citenamefont {Chekhova},\ and\ \citenamefont {Khalili}}]{knyazev2018quantum}%
  \BibitemOpen
  \bibfield  {author} {\bibinfo {author} {\bibfnamefont {E.}~\bibnamefont {Knyazev}}, \bibinfo {author} {\bibfnamefont {K.~Y.}\ \bibnamefont {Spasibko}}, \bibinfo {author} {\bibfnamefont {M.~V.}\ \bibnamefont {Chekhova}},\ and\ \bibinfo {author} {\bibfnamefont {F.~Y.}\ \bibnamefont {Khalili}},\ }\bibfield  {title} {\bibinfo {title} {Quantum tomography enhanced through parametric amplification},\ }\href {https://iopscience.iop.org/article/10.1088/1367-2630/aa99b4} {\bibfield  {journal} {\bibinfo  {journal} {New Journal of Physics}\ }\textbf {\bibinfo {volume} {20}},\ \bibinfo {pages} {013005} (\bibinfo {year} {2018})}\BibitemShut {NoStop}%
\bibitem [{\citenamefont {Yoon}\ \emph {et~al.}(2026)\citenamefont {Yoon}, \citenamefont {Roh}, \citenamefont {Gwak},\ and\ \citenamefont {Ra}}]{yoon2026efficient}%
  \BibitemOpen
  \bibfield  {author} {\bibinfo {author} {\bibfnamefont {Y.-D.}\ \bibnamefont {Yoon}}, \bibinfo {author} {\bibfnamefont {C.}~\bibnamefont {Roh}}, \bibinfo {author} {\bibfnamefont {G.}~\bibnamefont {Gwak}},\ and\ \bibinfo {author} {\bibfnamefont {Y.-S.}\ \bibnamefont {Ra}},\ }\bibfield  {title} {\bibinfo {title} {Efficient ultrafast homodyne detection of quantum light},\ }\href {https://arxiv.org/abs/2605.14858} {\bibfield  {journal} {\bibinfo  {journal} {arXiv:2605.14858}\ } (\bibinfo {year} {2026})}\BibitemShut {NoStop}%
\bibitem [{\citenamefont {Stammer}\ \emph {et~al.}(2025{\natexlab{a}})\citenamefont {Stammer}, \citenamefont {Rivera-Dean}, \citenamefont {Tzallas}, \citenamefont {Ciappina},\ and\ \citenamefont {Lewenstein}}]{stammer2025colloquium}%
  \BibitemOpen
  \bibfield  {author} {\bibinfo {author} {\bibfnamefont {P.}~\bibnamefont {Stammer}}, \bibinfo {author} {\bibfnamefont {J.}~\bibnamefont {Rivera-Dean}}, \bibinfo {author} {\bibfnamefont {P.}~\bibnamefont {Tzallas}}, \bibinfo {author} {\bibfnamefont {M.~F.}\ \bibnamefont {Ciappina}},\ and\ \bibinfo {author} {\bibfnamefont {M.}~\bibnamefont {Lewenstein}},\ }\bibfield  {title} {\bibinfo {title} {Colloquium: Quantum optics of intense light--matter interaction},\ }\href {https://arxiv.org/abs/2510.19045} {\bibfield  {journal} {\bibinfo  {journal} {arXiv:2510.19045}\ } (\bibinfo {year} {2025}{\natexlab{a}})}\BibitemShut {NoStop}%
\bibitem [{\citenamefont {Spasibko}\ \emph {et~al.}(2017)\citenamefont {Spasibko}, \citenamefont {Kopylov}, \citenamefont {Krutyanskiy}, \citenamefont {Murzina}, \citenamefont {Leuchs},\ and\ \citenamefont {Chekhova}}]{spasibko2017multiphoton}%
  \BibitemOpen
  \bibfield  {author} {\bibinfo {author} {\bibfnamefont {K.~Y.}\ \bibnamefont {Spasibko}}, \bibinfo {author} {\bibfnamefont {D.~A.}\ \bibnamefont {Kopylov}}, \bibinfo {author} {\bibfnamefont {V.~L.}\ \bibnamefont {Krutyanskiy}}, \bibinfo {author} {\bibfnamefont {T.~V.}\ \bibnamefont {Murzina}}, \bibinfo {author} {\bibfnamefont {G.}~\bibnamefont {Leuchs}},\ and\ \bibinfo {author} {\bibfnamefont {M.~V.}\ \bibnamefont {Chekhova}},\ }\bibfield  {title} {\bibinfo {title} {Multiphoton {Effects} {Enhanced} due to {Ultrafast} {Photon}-{Number} {Fluctuations}},\ }\href {https://doi.org/10.1103/PhysRevLett.119.223603} {\bibfield  {journal} {\bibinfo  {journal} {Phys. Rev. Lett.}\ }\textbf {\bibinfo {volume} {119}},\ \bibinfo {pages} {223603} (\bibinfo {year} {2017})}\BibitemShut {NoStop}%
\bibitem [{\citenamefont {Rasputnyi}\ \emph {et~al.}(2024)\citenamefont {Rasputnyi}, \citenamefont {Chen}, \citenamefont {Birk}, \citenamefont {Cohen}, \citenamefont {Kaminer}, \citenamefont {Kr{\"u}ger}, \citenamefont {Seletskiy}, \citenamefont {Chekhova},\ and\ \citenamefont {Tani}}]{rasputnyi2024high}%
  \BibitemOpen
  \bibfield  {author} {\bibinfo {author} {\bibfnamefont {A.}~\bibnamefont {Rasputnyi}}, \bibinfo {author} {\bibfnamefont {Z.}~\bibnamefont {Chen}}, \bibinfo {author} {\bibfnamefont {M.}~\bibnamefont {Birk}}, \bibinfo {author} {\bibfnamefont {O.}~\bibnamefont {Cohen}}, \bibinfo {author} {\bibfnamefont {I.}~\bibnamefont {Kaminer}}, \bibinfo {author} {\bibfnamefont {M.}~\bibnamefont {Kr{\"u}ger}}, \bibinfo {author} {\bibfnamefont {D.}~\bibnamefont {Seletskiy}}, \bibinfo {author} {\bibfnamefont {M.}~\bibnamefont {Chekhova}},\ and\ \bibinfo {author} {\bibfnamefont {F.}~\bibnamefont {Tani}},\ }\bibfield  {title} {\bibinfo {title} {High-harmonic generation by a bright squeezed vacuum},\ }\href {https://www.nature.com/articles/s41567-024-02659-x} {\bibfield  {journal} {\bibinfo  {journal} {Nature Physics}\ }\textbf {\bibinfo {volume} {20}},\ \bibinfo {pages} {1960} (\bibinfo {year} {2024})}\BibitemShut {NoStop}%
\bibitem [{\citenamefont {Heimerl}\ \emph {et~al.}(2024)\citenamefont {Heimerl}, \citenamefont {Mikhaylov}, \citenamefont {Meier}, \citenamefont {Höllerer}, \citenamefont {Kaminer}, \citenamefont {Chekhova},\ and\ \citenamefont {Hommelhoff}}]{heimerl_multiphoton_2024}%
  \BibitemOpen
  \bibfield  {author} {\bibinfo {author} {\bibfnamefont {J.}~\bibnamefont {Heimerl}}, \bibinfo {author} {\bibfnamefont {A.}~\bibnamefont {Mikhaylov}}, \bibinfo {author} {\bibfnamefont {S.}~\bibnamefont {Meier}}, \bibinfo {author} {\bibfnamefont {H.}~\bibnamefont {Höllerer}}, \bibinfo {author} {\bibfnamefont {I.}~\bibnamefont {Kaminer}}, \bibinfo {author} {\bibfnamefont {M.}~\bibnamefont {Chekhova}},\ and\ \bibinfo {author} {\bibfnamefont {P.}~\bibnamefont {Hommelhoff}},\ }\bibfield  {title} {\bibinfo {title} {Multiphoton electron emission with non-classical light},\ }\href {https://doi.org/10.1038/s41567-024-02472-6} {\bibfield  {journal} {\bibinfo  {journal} {Nat. Phys.}\ }\textbf {\bibinfo {volume} {20}},\ \bibinfo {pages} {945} (\bibinfo {year} {2024})}\BibitemShut {NoStop}%
\bibitem [{\citenamefont {Heimerl}\ \emph {et~al.}(2025)\citenamefont {Heimerl}, \citenamefont {Rasputnyi}, \citenamefont {P{\"o}lloth}, \citenamefont {Meier}, \citenamefont {Chekhova},\ and\ \citenamefont {Hommelhoff}}]{heimerl2025quantum}%
  \BibitemOpen
  \bibfield  {author} {\bibinfo {author} {\bibfnamefont {J.}~\bibnamefont {Heimerl}}, \bibinfo {author} {\bibfnamefont {A.}~\bibnamefont {Rasputnyi}}, \bibinfo {author} {\bibfnamefont {J.}~\bibnamefont {P{\"o}lloth}}, \bibinfo {author} {\bibfnamefont {S.}~\bibnamefont {Meier}}, \bibinfo {author} {\bibfnamefont {M.}~\bibnamefont {Chekhova}},\ and\ \bibinfo {author} {\bibfnamefont {P.}~\bibnamefont {Hommelhoff}},\ }\bibfield  {title} {\bibinfo {title} {Quantum light drives electrons strongly at metal needle tips},\ }\href {https://www.nature.com/articles/s41567-025-03087-1} {\bibfield  {journal} {\bibinfo  {journal} {Nature Physics}\ }\textbf {\bibinfo {volume} {21}},\ \bibinfo {pages} {1899} (\bibinfo {year} {2025})}\BibitemShut {NoStop}%
\bibitem [{\citenamefont {Lemieux}\ \emph {et~al.}(2025)\citenamefont {Lemieux}, \citenamefont {Jalil}, \citenamefont {Purschke}, \citenamefont {Boroumand}, \citenamefont {Hammond}, \citenamefont {Villeneuve}, \citenamefont {Naumov}, \citenamefont {Brabec},\ and\ \citenamefont {Vampa}}]{lemieux2024photon}%
  \BibitemOpen
  \bibfield  {author} {\bibinfo {author} {\bibfnamefont {S.}~\bibnamefont {Lemieux}}, \bibinfo {author} {\bibfnamefont {S.~A.}\ \bibnamefont {Jalil}}, \bibinfo {author} {\bibfnamefont {D.~N.}\ \bibnamefont {Purschke}}, \bibinfo {author} {\bibfnamefont {N.}~\bibnamefont {Boroumand}}, \bibinfo {author} {\bibfnamefont {T.}~\bibnamefont {Hammond}}, \bibinfo {author} {\bibfnamefont {D.}~\bibnamefont {Villeneuve}}, \bibinfo {author} {\bibfnamefont {A.}~\bibnamefont {Naumov}}, \bibinfo {author} {\bibfnamefont {T.}~\bibnamefont {Brabec}},\ and\ \bibinfo {author} {\bibfnamefont {G.}~\bibnamefont {Vampa}},\ }\bibfield  {title} {\bibinfo {title} {Photon bunching in high-harmonic emission controlled by quantum light},\ }\href {https://www.nature.com/articles/s41566-025-01673-6} {\bibfield  {journal} {\bibinfo  {journal} {Nature Photonics}\ }\textbf {\bibinfo {volume} {19}},\ \bibinfo {pages} {767} (\bibinfo {year} {2025})}\BibitemShut {NoStop}%
\bibitem [{\citenamefont {Kern}\ \emph {et~al.}(2026)\citenamefont {Kern}, \citenamefont {Nisim}, \citenamefont {Birk}, \citenamefont {Rasputnyi}, \citenamefont {Behar}, \citenamefont {Chen}, \citenamefont {Kaminer}, \citenamefont {Sidorenko}, \citenamefont {Cohen},\ and\ \citenamefont {Kr{\"u}ger}}]{kern2026single}%
  \BibitemOpen
  \bibfield  {author} {\bibinfo {author} {\bibfnamefont {Y.}~\bibnamefont {Kern}}, \bibinfo {author} {\bibfnamefont {I.}~\bibnamefont {Nisim}}, \bibinfo {author} {\bibfnamefont {M.}~\bibnamefont {Birk}}, \bibinfo {author} {\bibfnamefont {A.}~\bibnamefont {Rasputnyi}}, \bibinfo {author} {\bibfnamefont {D.}~\bibnamefont {Behar}}, \bibinfo {author} {\bibfnamefont {Z.}~\bibnamefont {Chen}}, \bibinfo {author} {\bibfnamefont {I.}~\bibnamefont {Kaminer}}, \bibinfo {author} {\bibfnamefont {P.}~\bibnamefont {Sidorenko}}, \bibinfo {author} {\bibfnamefont {O.}~\bibnamefont {Cohen}},\ and\ \bibinfo {author} {\bibfnamefont {M.}~\bibnamefont {Kr{\"u}ger}},\ }\bibfield  {title} {\bibinfo {title} {Single-shot pulse retrieval of femtosecond bright squeezed vacuum},\ }\href {https://doi.org/10.1364/OPTICA.580767} {\bibfield  {journal} {\bibinfo  {journal} {Optica}\ }\textbf {\bibinfo {volume} {13}},\ \bibinfo {pages} {395} (\bibinfo {year} {2026})}\BibitemShut {NoStop}%
\bibitem [{\citenamefont {Stammer}(2024)}]{stammer2024limitations}%
  \BibitemOpen
  \bibfield  {author} {\bibinfo {author} {\bibfnamefont {P.}~\bibnamefont {Stammer}},\ }\bibfield  {title} {\bibinfo {title} {On the limitations of the semi-classical picture in high harmonic generation},\ }\href {https://www.nature.com/articles/s41567-024-02579-w} {\bibfield  {journal} {\bibinfo  {journal} {Nature Physics}\ }\textbf {\bibinfo {volume} {20}},\ \bibinfo {pages} {1040} (\bibinfo {year} {2024})}\BibitemShut {NoStop}%
\bibitem [{\citenamefont {Cruz-Rodriguez}\ \emph {et~al.}(2024)\citenamefont {Cruz-Rodriguez}, \citenamefont {Dey}, \citenamefont {Freibert},\ and\ \citenamefont {Stammer}}]{cruz2024quantum}%
  \BibitemOpen
  \bibfield  {author} {\bibinfo {author} {\bibfnamefont {L.}~\bibnamefont {Cruz-Rodriguez}}, \bibinfo {author} {\bibfnamefont {D.}~\bibnamefont {Dey}}, \bibinfo {author} {\bibfnamefont {A.}~\bibnamefont {Freibert}},\ and\ \bibinfo {author} {\bibfnamefont {P.}~\bibnamefont {Stammer}},\ }\bibfield  {title} {\bibinfo {title} {Quantum phenomena in attosecond science},\ }\href {https://doi.org/10.1038/s42254-024-00769-2} {\bibfield  {journal} {\bibinfo  {journal} {Nat. Rev. Phys.}\ }\textbf {\bibinfo {volume} {6}},\ \bibinfo {pages} {691} (\bibinfo {year} {2024})}\BibitemShut {NoStop}%
\bibitem [{\citenamefont {Rivera-Dean}\ \emph {et~al.}(2025)\citenamefont {Rivera-Dean}, \citenamefont {Stammer}, \citenamefont {Ciappina},\ and\ \citenamefont {Lewenstein}}]{rivera2025structured}%
  \BibitemOpen
  \bibfield  {author} {\bibinfo {author} {\bibfnamefont {J.}~\bibnamefont {Rivera-Dean}}, \bibinfo {author} {\bibfnamefont {P.}~\bibnamefont {Stammer}}, \bibinfo {author} {\bibfnamefont {M.}~\bibnamefont {Ciappina}},\ and\ \bibinfo {author} {\bibfnamefont {M.}~\bibnamefont {Lewenstein}},\ }\bibfield  {title} {\bibinfo {title} {Structured squeezed light allows for high-harmonic generation in classical forbidden geometries},\ }\href {https://journals.aps.org/prl/abstract/10.1103/4hdl-bdwj} {\bibfield  {journal} {\bibinfo  {journal} {Phys. Rev. Lett.}\ }\textbf {\bibinfo {volume} {135}},\ \bibinfo {pages} {013801} (\bibinfo {year} {2025})}\BibitemShut {NoStop}%
\bibitem [{\citenamefont {Stammer}\ \emph {et~al.}(2026{\natexlab{a}})\citenamefont {Stammer}, \citenamefont {Granados},\ and\ \citenamefont {Rivera-Dean}}]{stammer2026fluctuation}%
  \BibitemOpen
  \bibfield  {author} {\bibinfo {author} {\bibfnamefont {P.}~\bibnamefont {Stammer}}, \bibinfo {author} {\bibfnamefont {C.}~\bibnamefont {Granados}},\ and\ \bibinfo {author} {\bibfnamefont {J.}~\bibnamefont {Rivera-Dean}},\ }\bibfield  {title} {\bibinfo {title} {Fluctuation-induced symmetry breaking in high harmonic generation for bicircular quantum light},\ }\href {https://arxiv.org/abs/2603.24377} {\bibfield  {journal} {\bibinfo  {journal} {arXiv:2603.24377}\ } (\bibinfo {year} {2026}{\natexlab{a}})}\BibitemShut {NoStop}%
\bibitem [{\citenamefont {Lewenstein}\ \emph {et~al.}(1994)\citenamefont {Lewenstein}, \citenamefont {Balcou}, \citenamefont {Ivanov}, \citenamefont {L’huillier},\ and\ \citenamefont {Corkum}}]{lewenstein1994theory}%
  \BibitemOpen
  \bibfield  {author} {\bibinfo {author} {\bibfnamefont {M.}~\bibnamefont {Lewenstein}}, \bibinfo {author} {\bibfnamefont {P.}~\bibnamefont {Balcou}}, \bibinfo {author} {\bibfnamefont {M.~Y.}\ \bibnamefont {Ivanov}}, \bibinfo {author} {\bibfnamefont {A.}~\bibnamefont {L’huillier}},\ and\ \bibinfo {author} {\bibfnamefont {P.~B.}\ \bibnamefont {Corkum}},\ }\bibfield  {title} {\bibinfo {title} {Theory of high-harmonic generation by low-frequency laser fields},\ }\href {https://link.aps.org/doi/10.1103/PhysRevA.49.2117} {\bibfield  {journal} {\bibinfo  {journal} {Physical Review A}\ }\textbf {\bibinfo {volume} {49}},\ \bibinfo {pages} {2117} (\bibinfo {year} {1994})}\BibitemShut {NoStop}%
\bibitem [{\citenamefont {Antoine}\ \emph {et~al.}(1996)\citenamefont {Antoine}, \citenamefont {L'huillier},\ and\ \citenamefont {Lewenstein}}]{antoine1996attosecond}%
  \BibitemOpen
  \bibfield  {author} {\bibinfo {author} {\bibfnamefont {P.}~\bibnamefont {Antoine}}, \bibinfo {author} {\bibfnamefont {A.}~\bibnamefont {L'huillier}},\ and\ \bibinfo {author} {\bibfnamefont {M.}~\bibnamefont {Lewenstein}},\ }\bibfield  {title} {\bibinfo {title} {Attosecond pulse trains using high--order harmonics},\ }\href {https://journals.aps.org/prl/abstract/10.1103/PhysRevLett.77.1234} {\bibfield  {journal} {\bibinfo  {journal} {Physical Review Letters}\ }\textbf {\bibinfo {volume} {77}},\ \bibinfo {pages} {1234} (\bibinfo {year} {1996})}\BibitemShut {NoStop}%
\bibitem [{\citenamefont {Lewenstein}\ \emph {et~al.}(2021)\citenamefont {Lewenstein}, \citenamefont {Ciappina}, \citenamefont {Pisanty}, \citenamefont {Rivera-Dean}, \citenamefont {Stammer}, \citenamefont {Lamprou},\ and\ \citenamefont {Tzallas}}]{lewenstein2021generation}%
  \BibitemOpen
  \bibfield  {author} {\bibinfo {author} {\bibfnamefont {M.}~\bibnamefont {Lewenstein}}, \bibinfo {author} {\bibfnamefont {M.~F.}\ \bibnamefont {Ciappina}}, \bibinfo {author} {\bibfnamefont {E.}~\bibnamefont {Pisanty}}, \bibinfo {author} {\bibfnamefont {J.}~\bibnamefont {Rivera-Dean}}, \bibinfo {author} {\bibfnamefont {P.}~\bibnamefont {Stammer}}, \bibinfo {author} {\bibfnamefont {T.}~\bibnamefont {Lamprou}},\ and\ \bibinfo {author} {\bibfnamefont {P.}~\bibnamefont {Tzallas}},\ }\bibfield  {title} {\bibinfo {title} {Generation of optical {S}chr{\"o}dinger cat states in intense laser--matter interactions},\ }\href {https://www.nature.com/articles/s41567-021-01317-w} {\bibfield  {journal} {\bibinfo  {journal} {Nature Physics}\ }\textbf {\bibinfo {volume} {17}},\ \bibinfo {pages} {1104} (\bibinfo {year} {2021})}\BibitemShut {NoStop}%
\bibitem [{\citenamefont {Gorlach}\ \emph {et~al.}(2020)\citenamefont {Gorlach}, \citenamefont {Neufeld}, \citenamefont {Rivera}, \citenamefont {Cohen},\ and\ \citenamefont {Kaminer}}]{gorlach2020quantum}%
  \BibitemOpen
  \bibfield  {author} {\bibinfo {author} {\bibfnamefont {A.}~\bibnamefont {Gorlach}}, \bibinfo {author} {\bibfnamefont {O.}~\bibnamefont {Neufeld}}, \bibinfo {author} {\bibfnamefont {N.}~\bibnamefont {Rivera}}, \bibinfo {author} {\bibfnamefont {O.}~\bibnamefont {Cohen}},\ and\ \bibinfo {author} {\bibfnamefont {I.}~\bibnamefont {Kaminer}},\ }\bibfield  {title} {\bibinfo {title} {The quantum-optical nature of high harmonic generation},\ }\href {https://www.nature.com/articles/s41467-020-18218-w} {\bibfield  {journal} {\bibinfo  {journal} {Nat. Commun.}\ }\textbf {\bibinfo {volume} {11}},\ \bibinfo {pages} {4598} (\bibinfo {year} {2020})}\BibitemShut {NoStop}%
\bibitem [{\citenamefont {Stammer}\ \emph {et~al.}(2024)\citenamefont {Stammer}, \citenamefont {Rivera-Dean}, \citenamefont {Maxwell}, \citenamefont {Lamprou}, \citenamefont {Arg{\"u}ello-Luengo}, \citenamefont {Tzallas}, \citenamefont {Ciappina},\ and\ \citenamefont {Lewenstein}}]{stammer2024entanglement}%
  \BibitemOpen
  \bibfield  {author} {\bibinfo {author} {\bibfnamefont {P.}~\bibnamefont {Stammer}}, \bibinfo {author} {\bibfnamefont {J.}~\bibnamefont {Rivera-Dean}}, \bibinfo {author} {\bibfnamefont {A.~S.}\ \bibnamefont {Maxwell}}, \bibinfo {author} {\bibfnamefont {T.}~\bibnamefont {Lamprou}}, \bibinfo {author} {\bibfnamefont {J.}~\bibnamefont {Arg{\"u}ello-Luengo}}, \bibinfo {author} {\bibfnamefont {P.}~\bibnamefont {Tzallas}}, \bibinfo {author} {\bibfnamefont {M.~F.}\ \bibnamefont {Ciappina}},\ and\ \bibinfo {author} {\bibfnamefont {M.}~\bibnamefont {Lewenstein}},\ }\bibfield  {title} {\bibinfo {title} {Entanglement and squeezing of the optical field modes in high harmonic generation},\ }\href {https://journals.aps.org/prl/abstract/10.1103/PhysRevLett.132.143603} {\bibfield  {journal} {\bibinfo  {journal} {Phys. Rev. Lett.}\ }\textbf {\bibinfo {volume} {132}},\ \bibinfo {pages} {143603} (\bibinfo {year} {2024})}\BibitemShut {NoStop}%
\bibitem [{\citenamefont {Yi}\ \emph {et~al.}(2025)\citenamefont {Yi}, \citenamefont {Klimkin}, \citenamefont {Brown}, \citenamefont {Smirnova}, \citenamefont {Patchkovskii}, \citenamefont {Babushkin},\ and\ \citenamefont {Ivanov}}]{yi2024generation}%
  \BibitemOpen
  \bibfield  {author} {\bibinfo {author} {\bibfnamefont {S.}~\bibnamefont {Yi}}, \bibinfo {author} {\bibfnamefont {N.~D.}\ \bibnamefont {Klimkin}}, \bibinfo {author} {\bibfnamefont {G.~G.}\ \bibnamefont {Brown}}, \bibinfo {author} {\bibfnamefont {O.}~\bibnamefont {Smirnova}}, \bibinfo {author} {\bibfnamefont {S.}~\bibnamefont {Patchkovskii}}, \bibinfo {author} {\bibfnamefont {I.}~\bibnamefont {Babushkin}},\ and\ \bibinfo {author} {\bibfnamefont {M.}~\bibnamefont {Ivanov}},\ }\bibfield  {title} {\bibinfo {title} {Generation of massively entangled bright states of light during harmonic generation in resonant media},\ }\href {https://link.aps.org/doi/10.1103/PhysRevX.15.011023} {\bibfield  {journal} {\bibinfo  {journal} {Physical Review X}\ }\textbf {\bibinfo {volume} {15}},\ \bibinfo {pages} {011023} (\bibinfo {year} {2025})}\BibitemShut {NoStop}%
\bibitem [{\citenamefont {Lange}\ \emph {et~al.}(2024)\citenamefont {Lange}, \citenamefont {Hansen},\ and\ \citenamefont {Madsen}}]{lange2024electron}%
  \BibitemOpen
  \bibfield  {author} {\bibinfo {author} {\bibfnamefont {C.~S.}\ \bibnamefont {Lange}}, \bibinfo {author} {\bibfnamefont {T.}~\bibnamefont {Hansen}},\ and\ \bibinfo {author} {\bibfnamefont {L.~B.}\ \bibnamefont {Madsen}},\ }\bibfield  {title} {\bibinfo {title} {Electron-correlation-induced nonclassicality of light from high-order harmonic generation},\ }\href {https://link.aps.org/doi/10.1103/PhysRevA.109.033110} {\bibfield  {journal} {\bibinfo  {journal} {Physical Review A}\ }\textbf {\bibinfo {volume} {109}},\ \bibinfo {pages} {033110} (\bibinfo {year} {2024})}\BibitemShut {NoStop}%
\bibitem [{\citenamefont {Lange}\ and\ \citenamefont {Madsen}(2025)}]{lange2024hierarchy}%
  \BibitemOpen
  \bibfield  {author} {\bibinfo {author} {\bibfnamefont {C.~S.}\ \bibnamefont {Lange}}\ and\ \bibinfo {author} {\bibfnamefont {L.~B.}\ \bibnamefont {Madsen}},\ }\bibfield  {title} {\bibinfo {title} {Hierarchy of approximations for describing quantum light from high-harmonic generation: A fermi-hubbard-model study},\ }\href {https://link.aps.org/doi/10.1103/PhysRevA.111.013113} {\bibfield  {journal} {\bibinfo  {journal} {Physical Review A}\ }\textbf {\bibinfo {volume} {111}},\ \bibinfo {pages} {013113} (\bibinfo {year} {2025})}\BibitemShut {NoStop}%
\bibitem [{\citenamefont {Stammer}\ \emph {et~al.}(2022)\citenamefont {Stammer}, \citenamefont {Rivera-Dean}, \citenamefont {Lamprou}, \citenamefont {Pisanty}, \citenamefont {Ciappina}, \citenamefont {Tzallas},\ and\ \citenamefont {Lewenstein}}]{stammer2022high}%
  \BibitemOpen
  \bibfield  {author} {\bibinfo {author} {\bibfnamefont {P.}~\bibnamefont {Stammer}}, \bibinfo {author} {\bibfnamefont {J.}~\bibnamefont {Rivera-Dean}}, \bibinfo {author} {\bibfnamefont {T.}~\bibnamefont {Lamprou}}, \bibinfo {author} {\bibfnamefont {E.}~\bibnamefont {Pisanty}}, \bibinfo {author} {\bibfnamefont {M.~F.}\ \bibnamefont {Ciappina}}, \bibinfo {author} {\bibfnamefont {P.}~\bibnamefont {Tzallas}},\ and\ \bibinfo {author} {\bibfnamefont {M.}~\bibnamefont {Lewenstein}},\ }\bibfield  {title} {\bibinfo {title} {High photon number entangled states and coherent state superposition from the extreme ultraviolet to the far infrared},\ }\href {https://link.aps.org/doi/10.1103/PhysRevLett.128.123603} {\bibfield  {journal} {\bibinfo  {journal} {Physical Review Letters}\ }\textbf {\bibinfo {volume} {128}},\ \bibinfo {pages} {123603} (\bibinfo {year} {2022})}\BibitemShut {NoStop}%
\bibitem [{\citenamefont {Gombk{\"o}t{\H{o}}}\ \emph {et~al.}(2021)\citenamefont {Gombk{\"o}t{\H{o}}}, \citenamefont {F{\"o}ldi},\ and\ \citenamefont {Varr{\'o}}}]{gombkotHo2021quantum}%
  \BibitemOpen
  \bibfield  {author} {\bibinfo {author} {\bibfnamefont {{\'A}.}~\bibnamefont {Gombk{\"o}t{\H{o}}}}, \bibinfo {author} {\bibfnamefont {P.}~\bibnamefont {F{\"o}ldi}},\ and\ \bibinfo {author} {\bibfnamefont {S.}~\bibnamefont {Varr{\'o}}},\ }\bibfield  {title} {\bibinfo {title} {Quantum-optical description of photon statistics and cross correlations in high-order harmonic generation},\ }\href {https://journals.aps.org/pra/abstract/10.1103/PhysRevA.104.033703} {\bibfield  {journal} {\bibinfo  {journal} {Phys. Rev. A}\ }\textbf {\bibinfo {volume} {104}},\ \bibinfo {pages} {033703} (\bibinfo {year} {2021})}\BibitemShut {NoStop}%
\bibitem [{\citenamefont {Stammer}\ \emph {et~al.}(2025{\natexlab{b}})\citenamefont {Stammer}, \citenamefont {Rivera-Dean},\ and\ \citenamefont {Lewenstein}}]{stammer2025theory}%
  \BibitemOpen
  \bibfield  {author} {\bibinfo {author} {\bibfnamefont {P.}~\bibnamefont {Stammer}}, \bibinfo {author} {\bibfnamefont {J.}~\bibnamefont {Rivera-Dean}},\ and\ \bibinfo {author} {\bibfnamefont {M.}~\bibnamefont {Lewenstein}},\ }\bibfield  {title} {\bibinfo {title} {Theory of quantum optics and optical coherence in high harmonic generation},\ }\href {https://arxiv.org/abs/2504.13287} {\bibfield  {journal} {\bibinfo  {journal} {arXiv:2504.13287}\ } (\bibinfo {year} {2025}{\natexlab{b}})}\BibitemShut {NoStop}%
\bibitem [{\citenamefont {Wang}\ and\ \citenamefont {Bian}(2025)}]{wang2025high}%
  \BibitemOpen
  \bibfield  {author} {\bibinfo {author} {\bibfnamefont {Y.-B.}\ \bibnamefont {Wang}}\ and\ \bibinfo {author} {\bibfnamefont {X.-B.}\ \bibnamefont {Bian}},\ }\bibfield  {title} {\bibinfo {title} {High-order harmonic generation in quantum light by a generalized von {Neumann} lattice method},\ }\href {https://journals.aps.org/pra/abstract/10.1103/PhysRevA.111.043111} {\bibfield  {journal} {\bibinfo  {journal} {Phys. Rev. A}\ }\textbf {\bibinfo {volume} {111}},\ \bibinfo {pages} {043111} (\bibinfo {year} {2025})}\BibitemShut {NoStop}%
\bibitem [{\citenamefont {Pizzi}\ \emph {et~al.}(2023)\citenamefont {Pizzi}, \citenamefont {Gorlach}, \citenamefont {Rivera}, \citenamefont {Nunnenkamp},\ and\ \citenamefont {Kaminer}}]{pizzi2023light}%
  \BibitemOpen
  \bibfield  {author} {\bibinfo {author} {\bibfnamefont {A.}~\bibnamefont {Pizzi}}, \bibinfo {author} {\bibfnamefont {A.}~\bibnamefont {Gorlach}}, \bibinfo {author} {\bibfnamefont {N.}~\bibnamefont {Rivera}}, \bibinfo {author} {\bibfnamefont {A.}~\bibnamefont {Nunnenkamp}},\ and\ \bibinfo {author} {\bibfnamefont {I.}~\bibnamefont {Kaminer}},\ }\bibfield  {title} {\bibinfo {title} {Light emission from strongly driven many-body systems},\ }\href {https://www.nature.com/articles/s41567-022-01910-7} {\bibfield  {journal} {\bibinfo  {journal} {Nature Physics}\ }\textbf {\bibinfo {volume} {19}},\ \bibinfo {pages} {551} (\bibinfo {year} {2023})}\BibitemShut {NoStop}%
\bibitem [{\citenamefont {Stammer}\ \emph {et~al.}(2026{\natexlab{b}})\citenamefont {Stammer}, \citenamefont {Rivera-Dean},\ and\ \citenamefont {Lewenstein}}]{stammer2026photon}%
  \BibitemOpen
  \bibfield  {author} {\bibinfo {author} {\bibfnamefont {P.}~\bibnamefont {Stammer}}, \bibinfo {author} {\bibfnamefont {J.}~\bibnamefont {Rivera-Dean}},\ and\ \bibinfo {author} {\bibfnamefont {M.}~\bibnamefont {Lewenstein}},\ }\bibfield  {title} {\bibinfo {title} {Photon anti-bunching in high harmonic generation},\ }\href {https://arxiv.org/abs/2606.17620} {\bibfield  {journal} {\bibinfo  {journal} {arXiv:2606.17620}\ } (\bibinfo {year} {2026}{\natexlab{b}})}\BibitemShut {NoStop}%
\bibitem [{\citenamefont {Lange}\ \emph {et~al.}(2025)\citenamefont {Lange}, \citenamefont {Hansen},\ and\ \citenamefont {Madsen}}]{lange_excitonic_2025}%
  \BibitemOpen
  \bibfield  {author} {\bibinfo {author} {\bibfnamefont {C.~S.}\ \bibnamefont {Lange}}, \bibinfo {author} {\bibfnamefont {T.}~\bibnamefont {Hansen}},\ and\ \bibinfo {author} {\bibfnamefont {L.~B.}\ \bibnamefont {Madsen}},\ }\bibfield  {title} {\bibinfo {title} {Excitonic {Enhancement} of {Squeezed} {Light} in {Quantum}-{Optical} {High}-{Harmonic} {Generation} from a {Mott} {Insulator}},\ }\href {https://doi.org/10.1103/wyk5-k8tk} {\bibfield  {journal} {\bibinfo  {journal} {Phys. Rev. Lett.}\ }\textbf {\bibinfo {volume} {135}},\ \bibinfo {pages} {043603} (\bibinfo {year} {2025})}\BibitemShut {NoStop}%
\bibitem [{\citenamefont {de-la Pe{\~n}a}\ \emph {et~al.}(2025)\citenamefont {de-la Pe{\~n}a}, \citenamefont {Neufeld}, \citenamefont {Even~Tzur}, \citenamefont {Cohen}, \citenamefont {Appel},\ and\ \citenamefont {Rubio}}]{de2024quantum}%
  \BibitemOpen
  \bibfield  {author} {\bibinfo {author} {\bibfnamefont {S.}~\bibnamefont {de-la Pe{\~n}a}}, \bibinfo {author} {\bibfnamefont {O.}~\bibnamefont {Neufeld}}, \bibinfo {author} {\bibfnamefont {M.}~\bibnamefont {Even~Tzur}}, \bibinfo {author} {\bibfnamefont {O.}~\bibnamefont {Cohen}}, \bibinfo {author} {\bibfnamefont {H.}~\bibnamefont {Appel}},\ and\ \bibinfo {author} {\bibfnamefont {A.}~\bibnamefont {Rubio}},\ }\bibfield  {title} {\bibinfo {title} {Quantum electrodynamics in high-harmonic generation: Multitrajectory ehrenfest and exact quantum analysis},\ }\href {https://doi.org/10.1021/acs.jctc.4c01206} {\bibfield  {journal} {\bibinfo  {journal} {J. Chem. Theory Comput.}\ }\textbf {\bibinfo {volume} {21}},\ \bibinfo {pages} {283} (\bibinfo {year} {2025})}\BibitemShut {NoStop}%
\bibitem [{\citenamefont {Boroumand}\ \emph {et~al.}(2025)\citenamefont {Boroumand}, \citenamefont {Thorpe}, \citenamefont {Bart}, \citenamefont {Wang}, \citenamefont {Purschke}, \citenamefont {Vampa},\ and\ \citenamefont {Brabec}}]{boroumand2025quantum}%
  \BibitemOpen
  \bibfield  {author} {\bibinfo {author} {\bibfnamefont {N.}~\bibnamefont {Boroumand}}, \bibinfo {author} {\bibfnamefont {A.}~\bibnamefont {Thorpe}}, \bibinfo {author} {\bibfnamefont {G.}~\bibnamefont {Bart}}, \bibinfo {author} {\bibfnamefont {L.}~\bibnamefont {Wang}}, \bibinfo {author} {\bibfnamefont {D.~N.}\ \bibnamefont {Purschke}}, \bibinfo {author} {\bibfnamefont {G.}~\bibnamefont {Vampa}},\ and\ \bibinfo {author} {\bibfnamefont {T.}~\bibnamefont {Brabec}},\ }\bibfield  {title} {\bibinfo {title} {Quantum engineering of high harmonic generation},\ }\href {https://arxiv.org/abs/2505.22536} {\bibfield  {journal} {\bibinfo  {journal} {arXiv:2505.22536}\ } (\bibinfo {year} {2025})}\BibitemShut {NoStop}%
\bibitem [{\citenamefont {Rivera-Dean}\ \emph {et~al.}(2026)\citenamefont {Rivera-Dean}, \citenamefont {Petrovic}, \citenamefont {Lewenstein},\ and\ \citenamefont {Stammer}}]{rivera2026attosecond}%
  \BibitemOpen
  \bibfield  {author} {\bibinfo {author} {\bibfnamefont {J.}~\bibnamefont {Rivera-Dean}}, \bibinfo {author} {\bibfnamefont {L.}~\bibnamefont {Petrovic}}, \bibinfo {author} {\bibfnamefont {M.}~\bibnamefont {Lewenstein}},\ and\ \bibinfo {author} {\bibfnamefont {P.}~\bibnamefont {Stammer}},\ }\bibfield  {title} {\bibinfo {title} {Attosecond quantum optical interferometry},\ }\href {https://iopscience.iop.org/article/10.1088/1361-6633/ae5847/meta} {\bibfield  {journal} {\bibinfo  {journal} {Reports on Progress in Physics}\ }\textbf {\bibinfo {volume} {89}},\ \bibinfo {pages} {047901} (\bibinfo {year} {2026})}\BibitemShut {NoStop}%
\bibitem [{\citenamefont {Stammer}\ \emph {et~al.}(2025{\natexlab{c}})\citenamefont {Stammer}, \citenamefont {Rivera-Dean}, \citenamefont {Ciappina},\ and\ \citenamefont {Lewenstein}}]{stammer2025weak}%
  \BibitemOpen
  \bibfield  {author} {\bibinfo {author} {\bibfnamefont {P.}~\bibnamefont {Stammer}}, \bibinfo {author} {\bibfnamefont {J.}~\bibnamefont {Rivera-Dean}}, \bibinfo {author} {\bibfnamefont {M.~F.}\ \bibnamefont {Ciappina}},\ and\ \bibinfo {author} {\bibfnamefont {M.}~\bibnamefont {Lewenstein}},\ }\bibfield  {title} {\bibinfo {title} {Weak measurement in strong laser field physics},\ }\href {https://arxiv.org/abs/2508.09048} {\bibfield  {journal} {\bibinfo  {journal} {arXiv:2508.09048}\ } (\bibinfo {year} {2025}{\natexlab{c}})}\BibitemShut {NoStop}%
\bibitem [{\citenamefont {Lamprou}\ \emph {et~al.}(2025)\citenamefont {Lamprou}, \citenamefont {Rivera-Dean}, \citenamefont {Stammer}, \citenamefont {Lewenstein},\ and\ \citenamefont {Tzallas}}]{lamprou2025nonlinear}%
  \BibitemOpen
  \bibfield  {author} {\bibinfo {author} {\bibfnamefont {T.}~\bibnamefont {Lamprou}}, \bibinfo {author} {\bibfnamefont {J.}~\bibnamefont {Rivera-Dean}}, \bibinfo {author} {\bibfnamefont {P.}~\bibnamefont {Stammer}}, \bibinfo {author} {\bibfnamefont {M.}~\bibnamefont {Lewenstein}},\ and\ \bibinfo {author} {\bibfnamefont {P.}~\bibnamefont {Tzallas}},\ }\bibfield  {title} {\bibinfo {title} {Nonlinear optics using intense optical coherent state superpositions},\ }\href {https://journals.aps.org/prl/abstract/10.1103/PhysRevLett.134.013601} {\bibfield  {journal} {\bibinfo  {journal} {Phys. Rev. Lett.}\ }\textbf {\bibinfo {volume} {134}},\ \bibinfo {pages} {013601} (\bibinfo {year} {2025})}\BibitemShut {NoStop}%
\bibitem [{\citenamefont {Dubois}\ \emph {et~al.}(2026)\citenamefont {Dubois}, \citenamefont {Cotte}, \citenamefont {Ta{\"\i}eb}, \citenamefont {L{\'e}v{\^e}que}, \citenamefont {Caillat}, \citenamefont {Dave}, \citenamefont {Sali{\`e}res}, \citenamefont {Bresteau}, \citenamefont {Bourassin-Bouchet}, \citenamefont {L'Huillier} \emph {et~al.}}]{dubois2026quantum}%
  \BibitemOpen
  \bibfield  {author} {\bibinfo {author} {\bibfnamefont {J.}~\bibnamefont {Dubois}}, \bibinfo {author} {\bibfnamefont {V.}~\bibnamefont {Cotte}}, \bibinfo {author} {\bibfnamefont {R.}~\bibnamefont {Ta{\"\i}eb}}, \bibinfo {author} {\bibfnamefont {C.}~\bibnamefont {L{\'e}v{\^e}que}}, \bibinfo {author} {\bibfnamefont {J.}~\bibnamefont {Caillat}}, \bibinfo {author} {\bibfnamefont {P.}~\bibnamefont {Dave}}, \bibinfo {author} {\bibfnamefont {P.}~\bibnamefont {Sali{\`e}res}}, \bibinfo {author} {\bibfnamefont {D.}~\bibnamefont {Bresteau}}, \bibinfo {author} {\bibfnamefont {C.}~\bibnamefont {Bourassin-Bouchet}}, \bibinfo {author} {\bibfnamefont {A.}~\bibnamefont {L'Huillier}}, \emph {et~al.},\ }\bibfield  {title} {\bibinfo {title} {Quantum optical photoelectron interferometry},\ }\href {https://arxiv.org/abs/2606.13447} {\bibfield  {journal} {\bibinfo  {journal} {arXiv:2606.13447}\ } (\bibinfo {year} {2026})}\BibitemShut {NoStop}%
\bibitem [{\citenamefont {Mele}\ \emph {et~al.}(2025)\citenamefont {Mele}, \citenamefont {Mele}, \citenamefont {Bittel}, \citenamefont {Eisert}, \citenamefont {Giovannetti}, \citenamefont {Lami}, \citenamefont {Leone},\ and\ \citenamefont {Oliviero}}]{mele2025learning}%
  \BibitemOpen
  \bibfield  {author} {\bibinfo {author} {\bibfnamefont {F.~A.}\ \bibnamefont {Mele}}, \bibinfo {author} {\bibfnamefont {A.~A.}\ \bibnamefont {Mele}}, \bibinfo {author} {\bibfnamefont {L.}~\bibnamefont {Bittel}}, \bibinfo {author} {\bibfnamefont {J.}~\bibnamefont {Eisert}}, \bibinfo {author} {\bibfnamefont {V.}~\bibnamefont {Giovannetti}}, \bibinfo {author} {\bibfnamefont {L.}~\bibnamefont {Lami}}, \bibinfo {author} {\bibfnamefont {L.}~\bibnamefont {Leone}},\ and\ \bibinfo {author} {\bibfnamefont {S.~F.}\ \bibnamefont {Oliviero}},\ }\bibfield  {title} {\bibinfo {title} {Learning quantum states of continuous-variable systems},\ }\href {https://www.nature.com/articles/s41567-025-03086-2} {\bibfield  {journal} {\bibinfo  {journal} {Nature Physics}\ ,\ \bibinfo {pages} {1}} (\bibinfo {year} {2025})}\BibitemShut {NoStop}%
\bibitem [{\citenamefont {Rasputnyi}\ \emph {et~al.}(2026)\citenamefont {Rasputnyi}, \citenamefont {Karuseichyk}, \citenamefont {Leuchs}, \citenamefont {Tani}, \citenamefont {Seletskiy},\ and\ \citenamefont {Chekhova}}]{rasputnyi2026kerr}%
  \BibitemOpen
  \bibfield  {author} {\bibinfo {author} {\bibfnamefont {A.}~\bibnamefont {Rasputnyi}}, \bibinfo {author} {\bibfnamefont {I.}~\bibnamefont {Karuseichyk}}, \bibinfo {author} {\bibfnamefont {G.}~\bibnamefont {Leuchs}}, \bibinfo {author} {\bibfnamefont {F.}~\bibnamefont {Tani}}, \bibinfo {author} {\bibfnamefont {D.}~\bibnamefont {Seletskiy}},\ and\ \bibinfo {author} {\bibfnamefont {M.}~\bibnamefont {Chekhova}},\ }\bibfield  {title} {\bibinfo {title} {Kerr-induced non-gaussianity of a bright ultrafast quantum state},\ }\href {https://doi.org/10.1364/OPTICA.599324} {\bibfield  {journal} {\bibinfo  {journal} {Optica}\ }\textbf {\bibinfo {volume} {13}},\ \bibinfo {pages} {1232} (\bibinfo {year} {2026})}\BibitemShut {NoStop}%
\bibitem [{\citenamefont {Yakovlev}\ \emph {et~al.}(2005)\citenamefont {Yakovlev}, \citenamefont {Bammer},\ and\ \citenamefont {Scrinzi}}]{yakovlev2005attosecond}%
  \BibitemOpen
  \bibfield  {author} {\bibinfo {author} {\bibfnamefont {V.~S.}\ \bibnamefont {Yakovlev}}, \bibinfo {author} {\bibfnamefont {F.}~\bibnamefont {Bammer}},\ and\ \bibinfo {author} {\bibfnamefont {A.}~\bibnamefont {Scrinzi}},\ }\bibfield  {title} {\bibinfo {title} {Attosecond streaking measurements},\ }\href {https://www.tandfonline.com/doi/full/10.1080/09500340412331283642} {\bibfield  {journal} {\bibinfo  {journal} {Journal of Modern Optics}\ }\textbf {\bibinfo {volume} {52}},\ \bibinfo {pages} {395} (\bibinfo {year} {2005})}\BibitemShut {NoStop}%
\bibitem [{\citenamefont {Yakovlev}\ \emph {et~al.}(2010)\citenamefont {Yakovlev}, \citenamefont {Gagnon}, \citenamefont {Karpowicz},\ and\ \citenamefont {Krausz}}]{yakovlev2010attosecond}%
  \BibitemOpen
  \bibfield  {author} {\bibinfo {author} {\bibfnamefont {V.~S.}\ \bibnamefont {Yakovlev}}, \bibinfo {author} {\bibfnamefont {J.}~\bibnamefont {Gagnon}}, \bibinfo {author} {\bibfnamefont {N.}~\bibnamefont {Karpowicz}},\ and\ \bibinfo {author} {\bibfnamefont {F.}~\bibnamefont {Krausz}},\ }\bibfield  {title} {\bibinfo {title} {Attosecond streaking enables the measurement of quantum phase},\ }\href {https://journals.aps.org/prl/abstract/10.1103/PhysRevLett.105.073001} {\bibfield  {journal} {\bibinfo  {journal} {Physical Review Letters}\ }\textbf {\bibinfo {volume} {105}},\ \bibinfo {pages} {073001} (\bibinfo {year} {2010})}\BibitemShut {NoStop}%
\bibitem [{\citenamefont {Stammer}\ \emph {et~al.}(2023)\citenamefont {Stammer}, \citenamefont {Rivera-Dean}, \citenamefont {Maxwell}, \citenamefont {Lamprou}, \citenamefont {Ord{\'o}{\~n}ez}, \citenamefont {Ciappina}, \citenamefont {Tzallas},\ and\ \citenamefont {Lewenstein}}]{stammer2023quantum}%
  \BibitemOpen
  \bibfield  {author} {\bibinfo {author} {\bibfnamefont {P.}~\bibnamefont {Stammer}}, \bibinfo {author} {\bibfnamefont {J.}~\bibnamefont {Rivera-Dean}}, \bibinfo {author} {\bibfnamefont {A.}~\bibnamefont {Maxwell}}, \bibinfo {author} {\bibfnamefont {T.}~\bibnamefont {Lamprou}}, \bibinfo {author} {\bibfnamefont {A.}~\bibnamefont {Ord{\'o}{\~n}ez}}, \bibinfo {author} {\bibfnamefont {M.~F.}\ \bibnamefont {Ciappina}}, \bibinfo {author} {\bibfnamefont {P.}~\bibnamefont {Tzallas}},\ and\ \bibinfo {author} {\bibfnamefont {M.}~\bibnamefont {Lewenstein}},\ }\bibfield  {title} {\bibinfo {title} {Quantum electrodynamics of intense laser-matter interactions: a tool for quantum state engineering},\ }\href {https://link.aps.org/doi/10.1103/PRXQuantum.4.010201} {\bibfield  {journal} {\bibinfo  {journal} {PRX Quantum}\ }\textbf {\bibinfo {volume} {4}},\ \bibinfo {pages} {010201} (\bibinfo {year} {2023})}\BibitemShut {NoStop}%
\bibitem [{\citenamefont {Fabre}\ and\ \citenamefont {Treps}(2020)}]{fabre2020modes}%
  \BibitemOpen
  \bibfield  {author} {\bibinfo {author} {\bibfnamefont {C.}~\bibnamefont {Fabre}}\ and\ \bibinfo {author} {\bibfnamefont {N.}~\bibnamefont {Treps}},\ }\bibfield  {title} {\bibinfo {title} {Modes and states in quantum optics},\ }\href {https://journals.aps.org/rmp/abstract/10.1103/RevModPhys.92.035005} {\bibfield  {journal} {\bibinfo  {journal} {Reviews of Modern Physics}\ }\textbf {\bibinfo {volume} {92}},\ \bibinfo {pages} {035005} (\bibinfo {year} {2020})}\BibitemShut {NoStop}%
\bibitem [{\citenamefont {Drummond}\ and\ \citenamefont {Gardiner}(1980)}]{drummond1980generalised}%
  \BibitemOpen
  \bibfield  {author} {\bibinfo {author} {\bibfnamefont {P.~D.}\ \bibnamefont {Drummond}}\ and\ \bibinfo {author} {\bibfnamefont {C.~W.}\ \bibnamefont {Gardiner}},\ }\bibfield  {title} {\bibinfo {title} {Generalised {P}-representations in quantum optics},\ }\href {https://doi.org/10.1088/0305-4470/13/7/018} {\bibfield  {journal} {\bibinfo  {journal} {J. Phys. A}\ }\textbf {\bibinfo {volume} {13}},\ \bibinfo {pages} {2353} (\bibinfo {year} {1980})}\BibitemShut {NoStop}%
\bibitem [{\citenamefont {Gilchrist}\ \emph {et~al.}(1997)\citenamefont {Gilchrist}, \citenamefont {Gardiner},\ and\ \citenamefont {Drummond}}]{gilchrist1997positive}%
  \BibitemOpen
  \bibfield  {author} {\bibinfo {author} {\bibfnamefont {A.}~\bibnamefont {Gilchrist}}, \bibinfo {author} {\bibfnamefont {C.}~\bibnamefont {Gardiner}},\ and\ \bibinfo {author} {\bibfnamefont {P.}~\bibnamefont {Drummond}},\ }\bibfield  {title} {\bibinfo {title} {Positive p representation: Application and validity},\ }\href {https://journals.aps.org/pra/abstract/10.1103/PhysRevA.55.3014} {\bibfield  {journal} {\bibinfo  {journal} {Physical Review A}\ }\textbf {\bibinfo {volume} {55}},\ \bibinfo {pages} {3014} (\bibinfo {year} {1997})}\BibitemShut {NoStop}%
\bibitem [{\citenamefont {Kim}\ \emph {et~al.}(1989)\citenamefont {Kim}, \citenamefont {De~Oliveira},\ and\ \citenamefont {Knight}}]{kim1989properties}%
  \BibitemOpen
  \bibfield  {author} {\bibinfo {author} {\bibfnamefont {M.}~\bibnamefont {Kim}}, \bibinfo {author} {\bibfnamefont {F.}~\bibnamefont {De~Oliveira}},\ and\ \bibinfo {author} {\bibfnamefont {P.}~\bibnamefont {Knight}},\ }\bibfield  {title} {\bibinfo {title} {Properties of squeezed number states and squeezed thermal states},\ }\href {https://journals.aps.org/pra/abstract/10.1103/PhysRevA.40.2494} {\bibfield  {journal} {\bibinfo  {journal} {Physical Review A}\ }\textbf {\bibinfo {volume} {40}},\ \bibinfo {pages} {2494} (\bibinfo {year} {1989})}\BibitemShut {NoStop}%
\bibitem [{\citenamefont {Olsen}\ and\ \citenamefont {Bradley}(2009)}]{olsen2009numerical}%
  \BibitemOpen
  \bibfield  {author} {\bibinfo {author} {\bibfnamefont {M.}~\bibnamefont {Olsen}}\ and\ \bibinfo {author} {\bibfnamefont {A.}~\bibnamefont {Bradley}},\ }\bibfield  {title} {\bibinfo {title} {Numerical representation of quantum states in the positive-p and wigner representations},\ }\href {https://www.sciencedirect.com/science/article/abs/pii/S0030401809005963} {\bibfield  {journal} {\bibinfo  {journal} {Optics Communications}\ }\textbf {\bibinfo {volume} {282}},\ \bibinfo {pages} {3924} (\bibinfo {year} {2009})}\BibitemShut {NoStop}%
\bibitem [{\citenamefont {Gorlach}\ \emph {et~al.}(2023)\citenamefont {Gorlach}, \citenamefont {Tzur}, \citenamefont {Birk}, \citenamefont {Kr{\"u}ger}, \citenamefont {Rivera}, \citenamefont {Cohen},\ and\ \citenamefont {Kaminer}}]{gorlach2023high}%
  \BibitemOpen
  \bibfield  {author} {\bibinfo {author} {\bibfnamefont {A.}~\bibnamefont {Gorlach}}, \bibinfo {author} {\bibfnamefont {M.~E.}\ \bibnamefont {Tzur}}, \bibinfo {author} {\bibfnamefont {M.}~\bibnamefont {Birk}}, \bibinfo {author} {\bibfnamefont {M.}~\bibnamefont {Kr{\"u}ger}}, \bibinfo {author} {\bibfnamefont {N.}~\bibnamefont {Rivera}}, \bibinfo {author} {\bibfnamefont {O.}~\bibnamefont {Cohen}},\ and\ \bibinfo {author} {\bibfnamefont {I.}~\bibnamefont {Kaminer}},\ }\bibfield  {title} {\bibinfo {title} {High-harmonic generation driven by quantum light},\ }\href {https://www.nature.com/articles/s41567-023-02127-y} {\bibfield  {journal} {\bibinfo  {journal} {Nature Physics}\ }\textbf {\bibinfo {volume} {19}},\ \bibinfo {pages} {1689} (\bibinfo {year} {2023})}\BibitemShut {NoStop}%
\bibitem [{\citenamefont {Gothelf}\ \emph {et~al.}(2026)\citenamefont {Gothelf}, \citenamefont {Madsen},\ and\ \citenamefont {Lange}}]{gothelf2026limitations}%
  \BibitemOpen
  \bibfield  {author} {\bibinfo {author} {\bibfnamefont {R.~V.}\ \bibnamefont {Gothelf}}, \bibinfo {author} {\bibfnamefont {L.~B.}\ \bibnamefont {Madsen}},\ and\ \bibinfo {author} {\bibfnamefont {C.~S.}\ \bibnamefont {Lange}},\ }\bibfield  {title} {\bibinfo {title} {Limitations of an approximative phase-space description in strong-field quantum optics},\ }\href {https://journals.aps.org/pra/abstract/10.1103/cxff-kky6} {\bibfield  {journal} {\bibinfo  {journal} {Physical Review A}\ }\textbf {\bibinfo {volume} {113}},\ \bibinfo {pages} {063107} (\bibinfo {year} {2026})}\BibitemShut {NoStop}%
\end{thebibliography}%

\end{document}